\documentclass[fleqn]{aa}
\usepackage{graphicx}
\usepackage{amssymb}
\usepackage{amsmath}
\usepackage{natbib}
\usepackage{mathrsfs}
\usepackage{ulem}
\usepackage{txfonts}
\usepackage{lmodern}
\usepackage{booktabs}
\usepackage{multicol}
\usepackage{cases}
\usepackage{color}

\usepackage[switch]{lineno} 

\newcommand{\corot}{{\textsc{CoRoT}}}

\newcommand{\tess}{{\textsc{TESS}}}
\newcommand{\mesa}{{\textsc{MESA}}}
\newcommand{\kepler}{\textit{Kepler}}

\newcommand{\ind}[1]{_{\rm #1}}

\def\m2s2{\,m$^{2}$\,s$^{-2}$} 

\def\aov{\alpha\ind{ov}}

\newcommand{\vaisala}{Brunt-V\"ais\"al\"a}

\newcommand{\rhoc}{\rho_{\rm c}}

\newcommand{\dpun}{\Delta\Pi_1}
\newcommand{\dn}{\Delta\nu}

\newenvironment{itemize*}%
  {\begin{itemize}%
    \setlength{\itemsep}{1pt}%
    \setlength{\parskip}{1pt}}%
  {\end{itemize}}

\newcommand{\omg}{\langle\Omega\rangle_{\rm g}}
\newcommand{\omp}{\langle\Omega\rangle_{\rm p}}

\newcommand\T{\rule{0pt}{2.6ex}}
\newcommand\B{\rule[-1.2ex]{0pt}{0pt}}

\begin{document}
\title{Seismic signature of electron degeneracy in the core of red giants: hints for mass transfer between close red-giant companions}
\titlerunning{Seismic signature of electron degeneracy in the core of red giants}
\author{
S. Deheuvels\inst{1}
\and J. Ballot\inst{1}
\and C. Gehan\inst{2,3}
\and B. Mosser\inst{4}
}

\institute{IRAP, Universit\'e de Toulouse, CNRS, CNES, UPS, 31400 Toulouse, France
\and Max Planck Institut f\"ur Sonnensystemforschung, Justus-von-Liebig-Weg 3, 37077 G\"ottingen, Germany
\and Instituto de Astrof\'isica e Ci\^encias do Espa\c{c}o, Universidade do Porto, CAUP, Rua das Estrelas, 4150-762 Porto, Portugal
\and LESIA, Observatoire de Paris, Universit\'e PSL, CNRS, Sorbonne Universit\'e, Universit\'e de Paris, 92195 Meudon, France
}

\offprints{S. Deheuvels\\ \email{sebastien.deheuvels@irap.omp.eu}
}


\abstract{The detection of mixed modes in red giants with space missions \corot\ and \kepler\ has revealed their deep internal structure. These modes allow us to characterize the pattern of pressure modes (through the measurement of their asymptotic frequency separation $\dn$) and the pattern of gravity modes (through the determination of their asymptotic period spacing $\dpun$). It has been shown that red giant branch (RGB) stars regroup on a well-defined sequence in the $\dn$-$\dpun$ plane. Our first goal is to theoretically explain the features of this sequence and understand how it can be used to probe the interiors of red giants. Using a grid of red giant models computed with \mesa, we demonstrate that red giants join the $\dn$-$\dpun$ sequence whenever electron degeneracy becomes strong in the core. We argue that this can be used to estimate the central densities of these stars, and potentially to measure the amount of core overshooting during the main sequence part of the evolution. We also investigate a puzzling subsample of red giants that are located below the RGB sequence, in contradiction with stellar evolution models. After checking the measurements of the asymptotic period spacing for these stars, we show that they are mainly intermediate-mass red giants. This is doubly peculiar because these stars should have nondegenerate cores and they are expected to be located well above the RGB sequence. We show that these peculiarities are well accounted for if these stars result from the interaction between two low-mass ($M\lesssim2\,M_\odot$) close companions during the red giant branch phase. If the secondary component has already developed a degenerate core before mass transfer begins, it becomes an intermediate-mass giant with a degenerate core. The secondary star is then located below the degenerate sequence, which is in agreement with the observations.}


\maketitle

\section{Introduction \label{sect_intro}}

Red giants stochastically excite oscillations in the outer convective envelope, in a similar way as the Sun. Contrary to main sequence solar-like pulsators, they show nonradial oscillations of a mixed nature, resulting from the coupling between pressure (p) waves and gravity (g) waves. These mixed modes have a very high potential in terms of asteroseismic diagnostics because they are sensitive to the envelope properties through their p-mode behavior, and to the core structure through their g-mode nature. High-precision space photometry provided by space missions \corot\ (\citealt{baglin06}), \kepler\ (\citealt{borucki10}), and now \tess\ (\citealt{ricker14}) has led to the detection of mixed modes in tens of thousands of red giants and thus revealed the internal structure of these stars. This has led to major achievements, such as the unambiguous distinction between H-shell burning giants and core-He burning giants (\citealt{bedding11}, \citealt{mosser11}). Mixed modes have also allowed us to probe the internal rotation of red giants \citep[e.g., ][]{beck12,mosser12b,deheuvels12,deheuvels14,deheuvels20,gehan18}, thereby permitting novel observational constraints to progress in the longstanding problem of the transport of angular momentum in stellar interiors (e.g., \citealt{marques13}, \citealt{cantiello14}, \citealt{fuller19}).

Well before the detection of mixed modes, \cite{shibahashi79} proposed an asymptotic expression for their frequencies. This later provided the proper framework to measure the global seismic parameters of p- and g-modes using mixed mode frequencies (\citealt{mosser12a}, \citealt{mosser15}). In particular, one can infer the asymptotic large separation $\dn$ of p modes and the asymptotic period spacing $\dpun$ of dipolar g modes. \cite{vrard16} (hereafter V16) thus measured $\dpun$ for 6,100 \kepler\ red giants. Placing red giants in the plane $\dpun$ versus $\dn$ provides a wealth of information about their structure. The most striking feature of this representation is the very clear separation between stars on the first-ascent red giant branch (RGB stars) and giants that are already burning He in the core (clump stars). Another remarkable feature is the well-defined sequence over which RGB stars regroup in the $\dn$-$\dpun$ plane (see Fig.\,\ref{dn_dp_vrard}). \cite{mosser14} found that as subgiants evolve onto the RGB, their evolutionary paths eventually converge in the $\dn$-$\dpun$ diagram. This RGB sequence in the $\dn$-$\dpun$ plane was also characterized by V16, who reported that the location of this sequence depends on the stellar mass. Interestingly, a similar sequence in the $\dn$-$\dpun$ diagram was also found for RGB stars using stellar models (\citealt{stello13}). However, no dedicated study has been proposed to explain the main features of this sequence. For future purposes, we also note that V16 found red giants that are located below this RGB sequence. We show in this study that stars are not expected to be found in this region of the $\dn$-$\dpun$ plane. 

In this paper, we investigate the RGB sequence in the $\dn$-$\dpun$ plane in order to physically explain its main features. Our first goal is to provide keys to interpret the comparison between the observed sequence and the one predicted by stellar models, which has been only briefly addressed in previous studies. In Sect.\,\ref{sect_obs}, we summarize the main observational features of the RGB sequence in the $\dn$-$\dpun$ diagram. In Sect.\,\ref{sect_deg}, we clearly establish the link between electron degeneracy in the core and the RGB sequence. We then use approximations in order to physically account for the observed features of the RGB sequence and we give hints on the way it could be used to probe the interiors of red giants (Sect.\,\ref{sect_theory}). In a second part of this paper, we instigate the puzzling question of stars that are located below the RGB sequence, in contradiction with the predictions of stellar models. In Sect.\,\ref{sect_compare}, we check the determination of $\dpun$ for a subset of this group of stars. Red giants that we confirm to be located below the RGB sequence are mainly intermediate-mass giants. This makes these stars doubly puzzling because they should have nondegenerate cores and are thus expected to have much larger period spacings, which would place them well above the RGB sequence. In Sect.\,\ref{sect_binary}, we show that these peculiarities could in fact be well accounted for if these stars are the product of mass transfer between two low-mass ($M\lesssim2\,M_\odot$) close companions during the red giant branch phase.

\section{Observational features of the RGB sequence in the $\dn$-$\dpun$ plane \label{sect_obs}}

\begin{figure}
\begin{center}
\includegraphics[width=9cm]{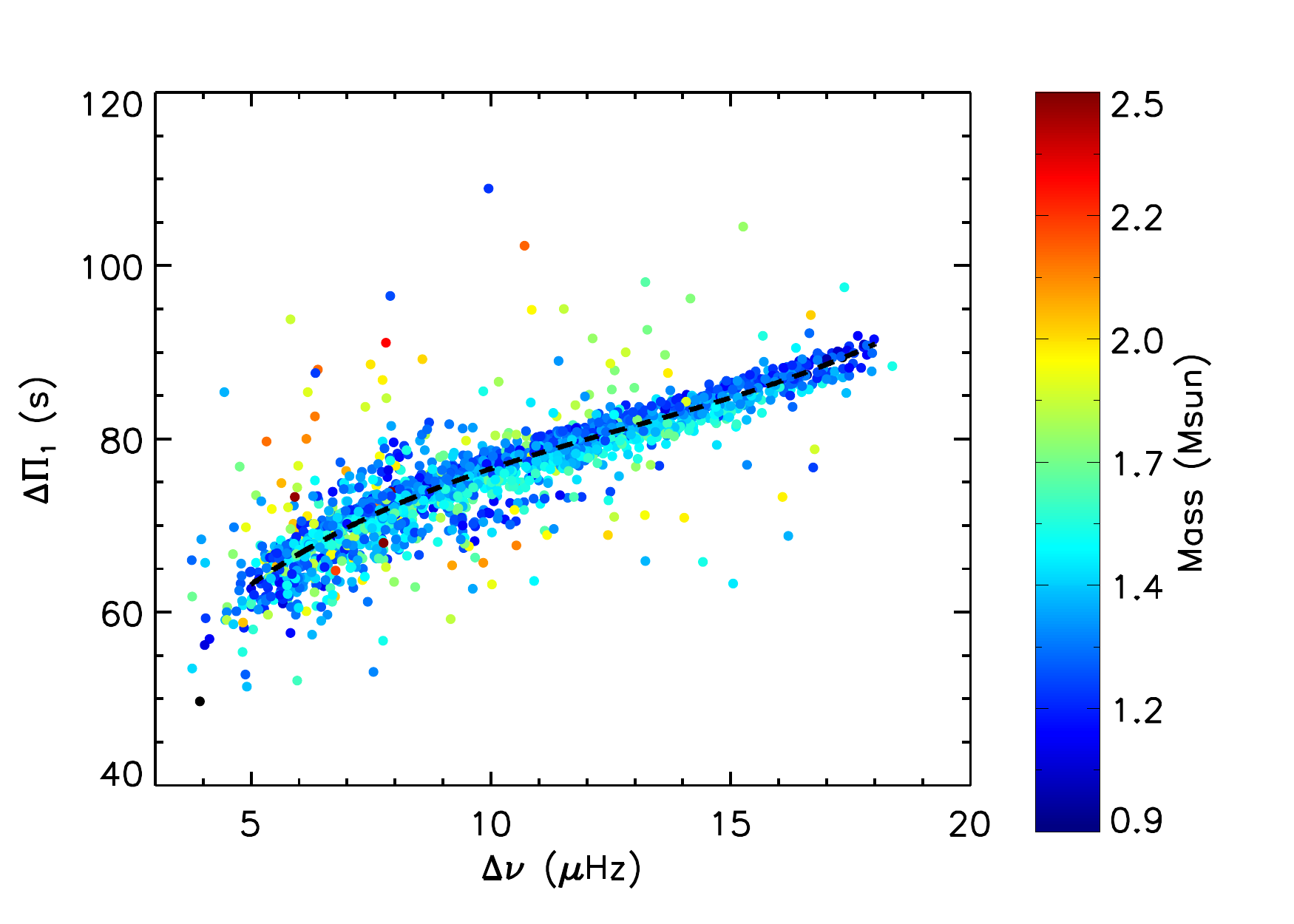}
\end{center}
\caption{$\Delta\nu$-$\Delta\Pi_1$ relation of RGB stars studied by V16. Only stars with $\Delta\Pi_1<130$ s are represented in order to remove the contribution from core-He burning giants. Stars are color-coded by their mass, obtained from seismic scaling relations. The thick black dashed line indicates the average RGB sequence as obtained from observations (see Sect.\,\ref{sect_z}).
\label{dn_dp_vrard}}
\end{figure}

The largest catalog of measured period spacings in red giants is that of V16, which contains 6 100 stars. The authors calculated the large separations $\Delta\nu$ using radial modes and an asymptotic expression of p modes (\citealt{mosser13}). The period spacings $\Delta\Pi_1$ were measured after applying a stretching of the observed mode periods in order to remove the dependency to p modes, as prescribed by \cite{mosser15}. Fig.\,\ref{dn_dp_vrard} shows the location of RGB stars in the $\dn$-$\dpun$ plane. Core-He burning giants were removed by selecting only stars with $\Delta\Pi_1<130$ s and we were thus left with close to 2000 stars on the RGB. RGB stars regroup on a well-defined sequence, whereby the period spacing decreases with decreasing $\Delta\nu$. As stars ascend the RGB, their radius increases and their large separation decreases, so that they evolve from right to left on this sequence. In Fig.\,\ref{dn_dp_vrard}, stars are color-coded by their masses, which were derived from seismic scaling relations by V16.

Several key features of the RGB sequence have already been observed. First, there is a clear mass-dependency of the RGB sequence. At fixed $\Delta\Pi_1$, higher-mass stars tend to have larger values of $\Delta\nu$, which means that they are denser than their lower-mass counterparts. This was pointed out by V16, but no physical explanation of this tendency has yet been proposed. Secondly, higher-mass stars appear to join the RGB sequence at lower $\Delta\nu$ (i.e., later in the evolution) than lower-mass stars. This is evident from Fig.\,\ref{dn_dp_vrard}, where the right part of the sequence is comprised only of low-mass stars. This was also found in stellar models (\citealt{lagarde16}). Finally, there is a scatter around the RGB sequence, which has been reported to increase with stellar mass (\citealt{mosser14}). At least part of this scatter could be caused by incorrect estimations of the period spacing. We propose physical explanations for these features in Sect.\,\ref{sect_deg} and \ref{sect_theory}.

\section{Link with electron degeneracy in the core \label{sect_deg}}

We know that the asymptotic large separation $\Delta\nu$ essentially measures the mean density of the star. The period spacing $\Delta\Pi_1$ is related to the properties of the stellar core. The existence of a tight sequence for RGB stars in the $\dn$-$\dpun$ diagram shows that stars with a given $\Delta\Pi_1$ and a given mass all have similar mean densities, that is, similar envelope properties. We here study the link between electron degeneracy in the core and the settling of stars on the RGB sequence in the $\dn$-$\dpun$ diagram, which was also suggested by \cite{farnir21}.

For this purpose, we computed a grid of stellar models using the evolution code \mesa\ \citep{paxton11,paxton13,paxton15,paxton18,paxton19}. This grid was further used to explore the dependence of the RGB sequence on different stellar parameters, like the mass, the metallicity and the amount of core overshoot during the main sequence. We thus varied the stellar mass from 1 to 2.5 $M_\odot$ (step 0.1 $M_\odot$). We considered a range of metallicity between $-0.4$ to 0.4 dex (step 0.1 dex), which covers the great majority of \kepler\ giants. We assumed the solar mixture of \cite{asplund09} for all our models. Nuclear reactions rates were computed with the NACRE compilation (\citealt{angulo99}) completed with the revised LUNA rate (\citealt{formicola04}) for the $^{14}$N$(p,\gamma)^{15}$O reaction. The atmosphere was described by Eddington's gray law. Convection was described using the classical mixing length theory (\citealt{bohm58}) calibrated on the Sun. Core overshooting was added during the main sequence as an instantaneous mixing beyond the convective core over a distance $d_{\rm ov} = \alpha_{\rm ov} H_P$, where $H_P$ is the  pressure scale height and $\alpha_{\rm ov}$ is a free parameter controlling the efficiency of core overshooting. We considered values of $\alpha_{\rm ov} = 0$, 0.1, and 0.2, in line with the existing observational measurements of this free parameter (e.g., \citealt{deheuvels16}). 

Models were evolved until they reach a large separation of 2\,$\mu$Hz, that is, well below the lower limit of $\dn$ for which measurements of $\dpun$ can be obtained (see Fig. \ref{dn_dp_vrard}).  At each stage of the evolution, we estimated $\Delta\nu$ as
\begin{equation}
\Delta\nu = \Delta\nu_\odot \left(\frac{M}{M_\odot}\right)^{1/2} \left(\frac{R}{R_\odot}\right)^{-3/2} 
\end{equation}
where the solar large separation was taken as $\Delta\nu_\odot = 134.9 \,\mu$Hz (\citealt{kallinger10}). $\Delta\Pi_1$ was estimated using its asymptotic expression
\begin{equation}
\Delta\Pi_1 = \pi^2\sqrt{2} \left( \int_{\rm g} \frac{N_{\rm BV}}{r} \,\hbox{d}r \right)^{-1}
\label{eq_dp}
\end{equation}
where $N_{\rm BV}$ is the \vaisala\ frequency and the integration is performed over the extent of the g-mode cavity. 

For each model, the level of degeneracy of the electrons in the core is measured by the parameter $\psi$ defined as
\begin{equation}
e^\psi = \frac{h^3}{2m_{\rm u}(2\pi m_{\rm e}k_{\rm B})^{3/2}} \frac{\rho}{\mu_{\rm e}T^{3/2}}
\end{equation}
where $\rho$ is the gas density, $T$ is its temperature, $\mu_{\rm e}$ is the mean molecular weight per free electron, $h$ is the Planck constant, $k_{\rm B}$ is the Boltzmann constant, $m_{\rm u}$ is the atomic unit mass, and $m_{\rm e}$ is the mass of the electron. Values of $\psi\gg 1$ indicate a high level of electron degeneracy (\citealt{kippenhahn90}). We here arbitrarily considered that the core has reached strong electron degeneracy when $\psi(r=0) > 25$.

\begin{figure}
\begin{center}
\includegraphics[width=9cm]{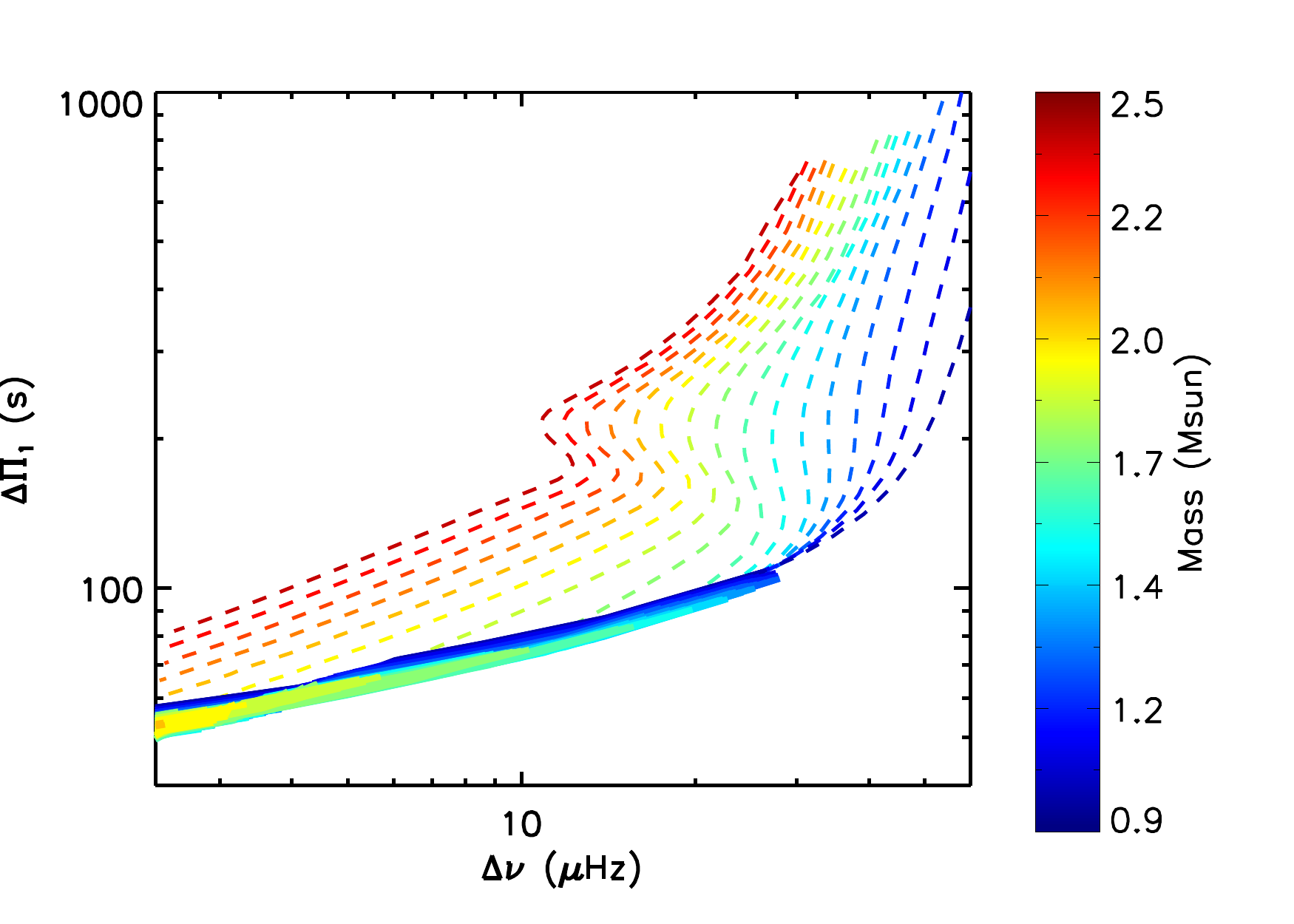}
\end{center}
\caption{Evolutionary paths in the $\dn$-$\dpun$ plane of stellar models from the grid presented in Sect.\,\ref{sect_deg}. Thick solid lines indicate periods of the evolution where electron degeneracy is strong in the core -- $\psi(r=0)>25$. Dashed lines correspond to $\psi(r=0)<25$. For clarity, the nondegenerate part of the evolution is shown only for models with solar metallicity and without core overshooting.
\label{dn_dp_deg}}
\end{figure}

Fig.\,\ref{dn_dp_deg} shows the evolutionary tracks of models from the grid. The parts of the evolution where electron degeneracy is high are indicated by the thick colored curves. For clarity reasons, we only represented the nondegenerate part of the evolution for models with solar metallicity and without core overshooting. It is very clear from this figure that red giants populating the RGB sequence in the $\dn$-$\dpun$ plane all have degenerate cores. Conversely, no models with a degenerate core are found outside this sequence. We thus propose to refer to this particular region in the $\dn$-$\dpun$ plane as the \textit{degenerate sequence}. 

In Fig.\,\ref{dn_dp_deg}, stars evolve from the top right corner to the bottom left corner. During the evolution, the core gets denser and therefore the asymptotic period spacing decreases. For stars that are less massive than about 2.1\, $M_\odot$, the core eventually reaches conditions of density and temperature such that electron degeneracy becomes important (thick colored curves in Fig.\,\ref{dn_dp_deg}). Then, $\dn$ and $\dpun$ vary following the degenerate sequence, which depends weakly on the stellar mass but seems nearly independent of other stellar parameters. It is well known that the higher the mass, the later in the evolution the He core becomes degenerate (e.g., \citealt{kippenhahn90}). This explains why higher-mass red giants join the sequence later than lower-mass red giants. 

Stars that are located above the degenerate sequence in the $\dn$-$\dpun$ have nondegenerate cores and are thus evolving on a Kelvin-Helmholtz timescale. This explains why we detect relatively few stars in this region of the diagram. On the other hand, no stars are expected to be located below the degenerate sequence. Yet Fig.\,\ref{dn_dp_vrard} shows that several stars are found in this part of the diagram. We address this intriguing question in Sect.\,\ref{sect_compare} and \ref{sect_binary}.

\section{Investigating the properties of the degenerate sequence \label{sect_theory}}

We used the grid of stellar models built in Sect.\,\ref{sect_deg} to investigate the observed features of the degenerate sequence. 

\subsection{Relation between $\Delta\Pi_1$ and $\rho_{\rm c}$ \label{sect_rhoc}}

\begin{figure}
\begin{center}
\includegraphics[width=9cm]{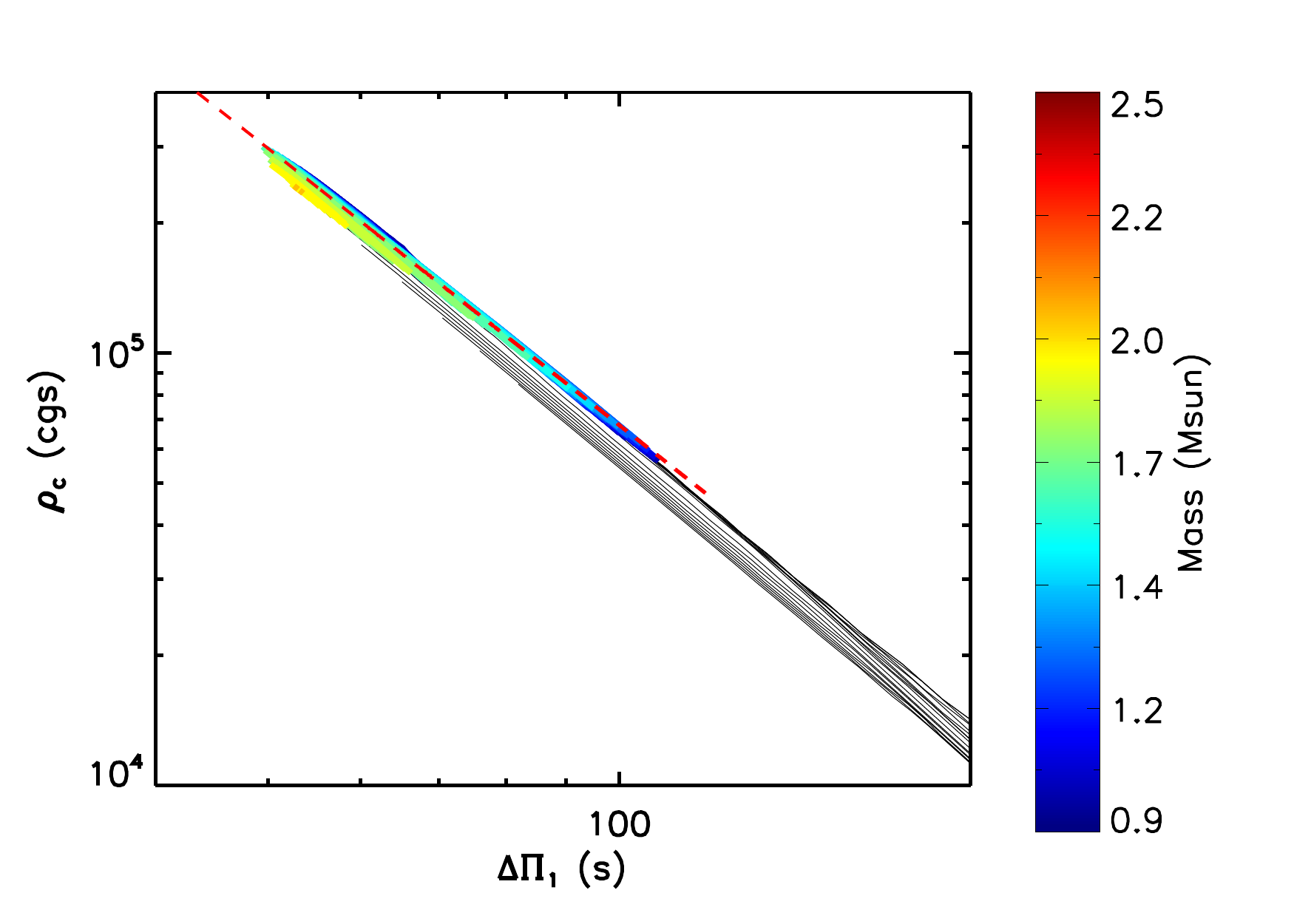}
\end{center}
\caption{Relation between $\Delta\Pi_1$ and the central density $\rho_{\rm c}$ in the models of the grid presented in Sect.\,\ref{sect_deg}. Thick colored lines (resp. thin black lines) indicate periods of the evolution where electron degeneracy is strong (resp. weak) in the core. The color code is the same as in Fig.\,\ref{dn_dp_deg}. The red dashed line corresponds to a linear regression of the $\Delta\Pi_1$-$\rho_{\rm c}$ relation for stars with degenerate cores (see text).
\label{dp_rhoc_deg}}
\end{figure}

%

It is quite well known that in subgiants and red giants, g-mode properties are tightly related to the central density $\rho_{\rm c}$ (e.g., \citealt{deheuvels11}, \citealt{montalban13}). Fig.\,\ref{dp_rhoc_deg} shows the relation between $\Delta\Pi_1$ and $\rho_{\rm c}$ for all the models of the computed grid. The correlation between the two quantities appears clearly, and becomes tighter as stars evolve (as $\Delta\Pi_1$ decreases). We observe that when the core becomes degenerate (thick colored curves in Fig.\,\ref{dp_rhoc_deg}), there is a nearly one-to-one relation between $\rho_{\rm c}$ and $\Delta\Pi_1$. For values of $\Delta\Pi_1$ above around 75 s, the $\rho_{\rm c}$-$\Delta\Pi_1$ relation is very tight. Despite the wide range of stellar parameters explored in our grid, we find that models sharing the same $\Delta\Pi_1$ within 0.5 s (typical uncertainty in the determination of $\Delta\Pi_1$, see V16) show variations in $\rhoc$ of only about 5\%. A large fraction of the RGB stars on the degenerate sequence in the V16 catalog are in this range of $\dpun$. For lower values of $\Delta\Pi_1$, the scatter in the relation is larger and reaches about 15\% for $\Delta\Pi_1\sim 50$ s. 

A first rough explanation of this tight relation can be obtained by assuming that the stellar matter in the core behaves as a completely degenerate nonrelativistic gas. This leads us to model the degenerate core as a polytrope with $P=K(\rho/\mu_{\rm e})^\gamma$, where $\gamma=5/3$, $\mu_{\rm e}$ is the mean molecular weight per free electron, and $K$ is a constant determined by the equation of state of the gas. Such models have one single degree of freedom. Their complete mechanical structure, that is $\rho(r)$, $P(r)$, $m(r)$, and thus $N_{\rm BV}(r)$ are entirely determined by one parameter only, for instance the central density $\rho_{\rm c}$. 

\begin{figure}
\begin{center}
\includegraphics[width=9cm]{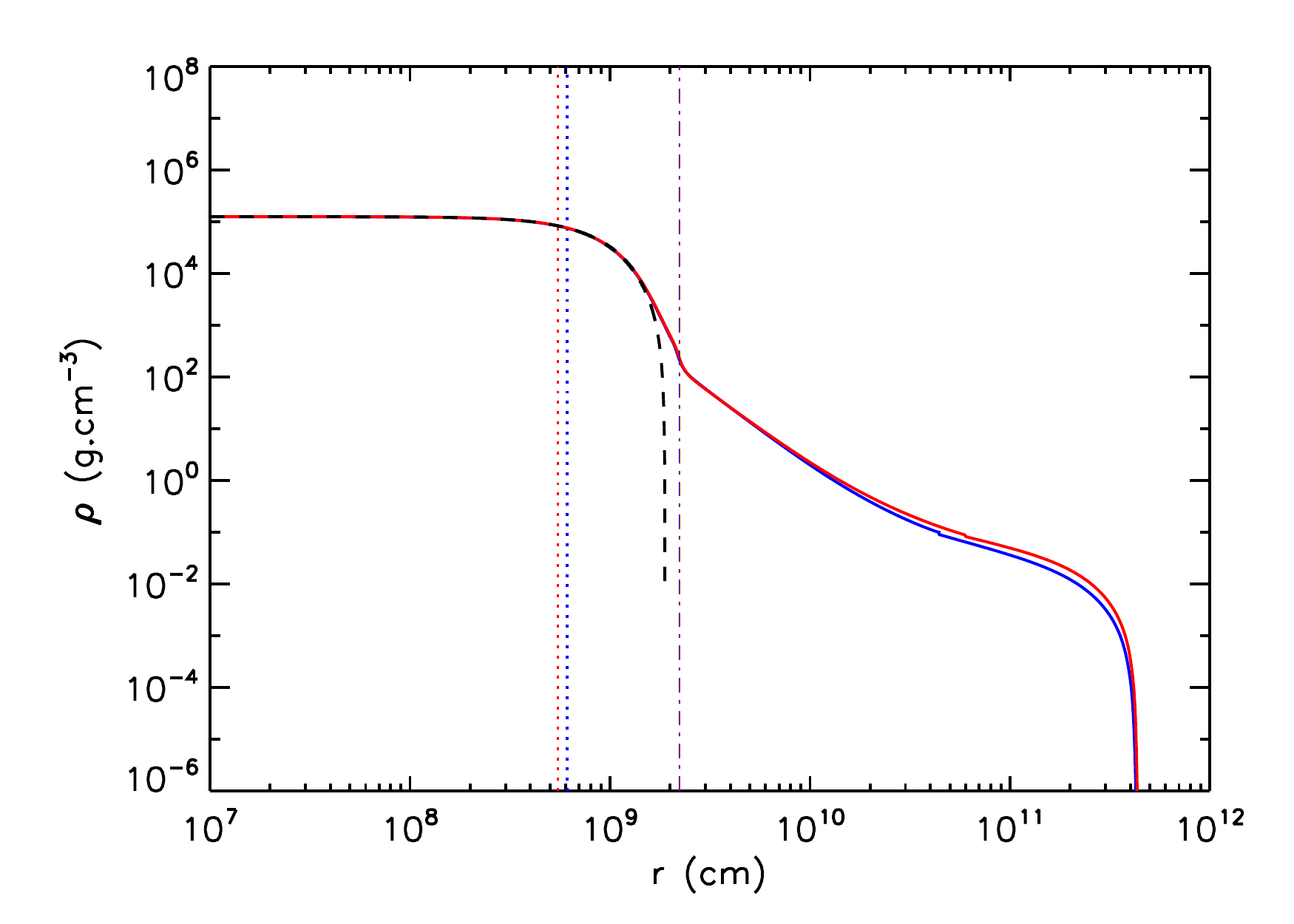}
\end{center}
\caption{Density profile in the degenerate helium core of models sharing similar values of $\rho_{\rm c}$ but different stellar parameters (blue curve: $M=1\,M_\odot$ and $[$Fe/H$] = -0.4$ dex, red curve: $M=1.4\,M_\odot$ and $[$Fe/H$] = +0.4$ dex). The dotted lines delimit the region of strong degeneracy for both models and the dash-dot lines indicate the location of the H-burning shell. The black dashed line corresponds to the density profile for a polytrope of exponent $\gamma = 1.55$.
\label{fig_core_deg}}
\end{figure}

In reality, the helium core is not fully degenerate and we do not expect the central layers to behave as a polytrope of exponent $\gamma=5/3$. To verify this, we calculated the local polytropic exponent $\gamma=(\hbox{d}\ln p/\hbox{d}\ln\rho)$ in the degenerate cores of the models in our computed grid. We found that $\gamma$ is nearly identical for models with degenerate cores, with values ranging from about 1.54 to about 1.56. We can thus make the assumption that the degenerate cores of models having similar central densities behave as polytropes with an identical exponent $\gamma$. To illustrate this, we compare in Fig.\,\ref{fig_core_deg} the density profiles of two models with different stellar parameters but same $\rho_{\rm c}$ to the density profile of a polytrope with $\gamma = 1.55$ and identical $\rho_{\rm c}$. The curves overlap in the region of strong degeneracy.


Assuming that the degenerate cores of RGB stars can be modeled as polytropes with the same index $n$, the expression of the \vaisala\ frequency can be recast as
\begin{equation}
N_{\rm BV}^2 = 4\pi G \rho_{\rm c} n \frac{1}{w}\left(\frac{\hbox{d}w}{\hbox{d}z}\right)^2 \left( 1 - \frac{\gamma}{\Gamma_1} \right),
\label{eq_bv_deg}
\end{equation}
where $G$ is the gravitational constant, $\Gamma_1$ is the adiabatic exponent $(\partial\ln P/\partial\ln\rho)_{\rm ad}$, $w=(\rho/\rhoc)^{1/n}$, and $z=r/A$, where
\begin{equation}
A = \left[ \frac{K(n+1)\rhoc^{1/n-1}}{4\pi G} \right]^{1/2}.
\end{equation}
The only free parameter governing the value of $N_{\rm BV}(r)$ is $\rhoc$, and we deduce that $N_{\rm BV} \propto \sqrt{\rhoc}$. 

The g-mode cavity extends up to the hydrogen burning shell and thus also comprises regions that are weakly degenerate. Eq. \ref{eq_bv_deg} cannot a priori hold in these regions. However, we observe from our grid that models sharing similar central densities also have very similar structure in the weakly-degenerate regions below the H-shell burning. This appears clearly in Fig.\,\ref{fig_core_deg}. We therefore assumed that the relation $N_{\rm BV} \sim \sqrt{\rhoc}$ holds in the whole g-mode cavity. In these conditions, Eq. \ref{eq_dp} shows that $\Delta\Pi_1 \sim \rhoc^{-1/2}$. To check how this relation agrees with the predictions from stellar models, we fit a power law to the $\Delta\Pi_1$-$\rhoc$ relation obtained from degenerate models in our grid (see Fig.\,\ref{dp_rhoc_deg}). We found $\Delta\Pi_1 \sim \rhoc^{-0.47}$, which is close to the dependence that was obtained with our simplifying assumptions. We conclude that for red giants belonging to the degenerate sequence, the asymptotic period spacing $\Delta\Pi_1$ provides a quite good direct estimate of the central density. 

\subsection{Relation between $\Delta\Pi_1$ and stellar radius \label{sect_dp_R}}

From the previous section, we can interpret the degenerate sequence in the $\dn$-$\dpun$ plane as a tight relation between the central density $\rho_{\rm c}$ (through $\Delta\Pi_1$) and the mean density $\bar{\rho}$ (through $\Delta\nu$). This is a manifestation of the more general mirror principle, whereby the envelope expands as the core contracts. In this particular case, the relation between core properties and envelope properties can be understood as follows. We have shown that stars with similar period spacings have nearly identical structure from the center to the H-burning shell. We can thus assume that the masses $M_{\rm c}$ and radii $R_{\rm c}$ of their inert helium cores are similar. From arguments based on shell-source homology (\citealt{refsdal70}), the stellar luminosity essentially depends on $M_{\rm c}$ and $R_{\rm c}$, so that stars with similar $\Delta\Pi_1$ should have approximately the same luminosity regardless of their total mass. The link with envelope properties can then be made by noticing that red giants have to evolve along Hayashi lines in the HR diagram. If we make the very crude approximation that all RGB stars follow the same Hayashi line, defined by $L(T_{\rm eff})$, then stars with similar $\Delta\Pi_1$ should have the same effective temperature, and thus the same radius (since $L\propto R^2 T_{\rm eff}^4$). We thus conclude that there should be a close relation between $\Delta\Pi_1$ and the stellar radius $R$.

\begin{figure}
\begin{center}
\includegraphics[width=9cm]{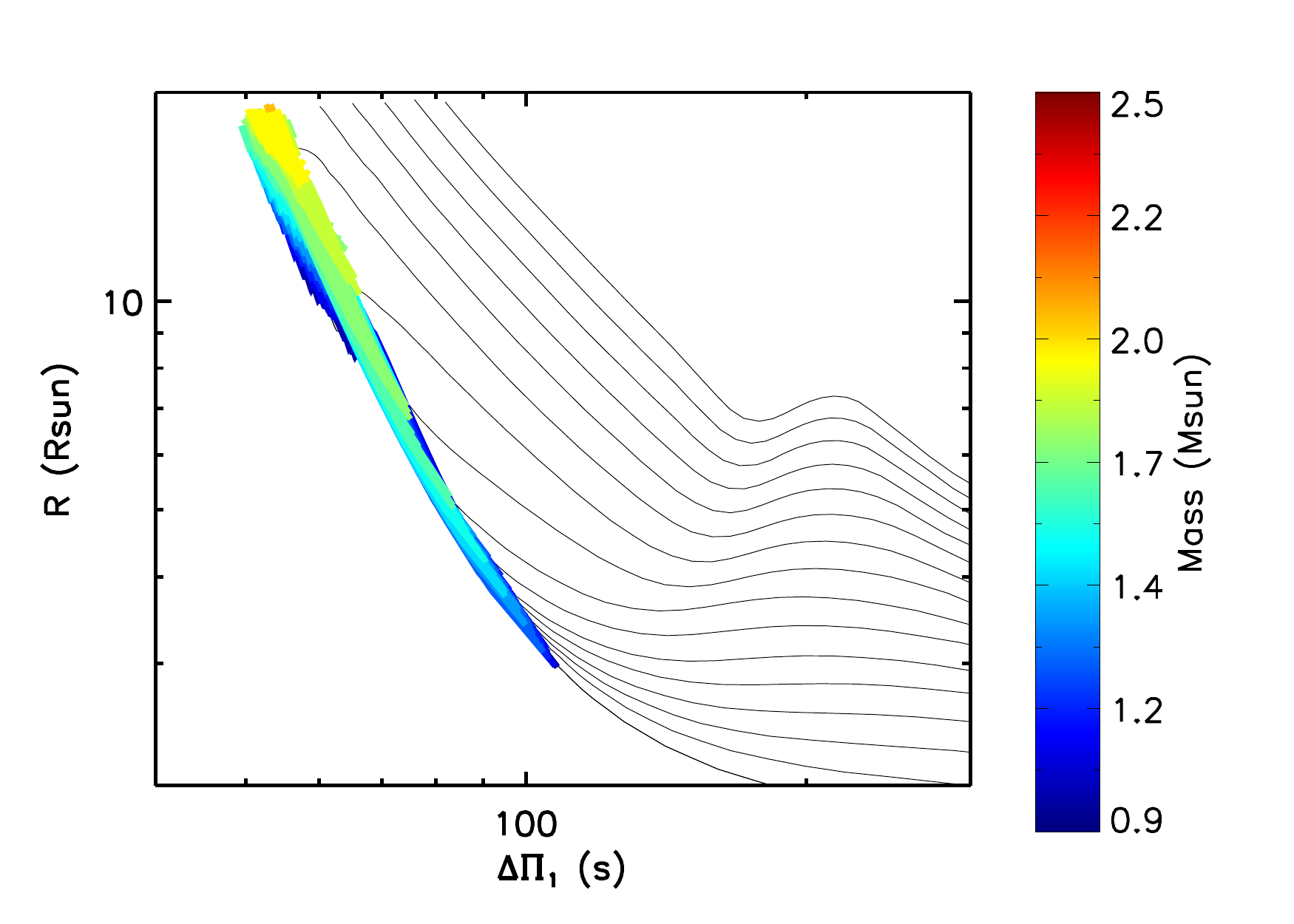}
\end{center}
\caption{Relation between $\Delta\Pi_1$ and the stellar radius $R$ in the models of the grid presented in Sect.\,\ref{sect_deg}. The colors and thickness of the lines have the same meaning as in Fig.\,\ref{dp_rhoc_deg}. For clarity, the nondegenerate part of the evolution is shown only for models with solar metallicity, $\aov = 0$, and masses ranging from 1 to 2.5\,$M_\odot$ from bottom to top. 
\label{rhoc_radius_deg}}
\end{figure}

As shown by Fig.\,\ref{rhoc_radius_deg}, we indeed find a tight relation between $\Delta\Pi_1$ and $R$ for the models of our grid that have degenerate cores. The relation is very clear for low-luminosity giants ($R \lesssim 8\,R_\odot$). Contrary to the relation between $\Delta\Pi_1$ and $\Delta\nu$, it seems to be nearly independent of the stellar mass. The widening of the relation above $\sim \, 8\,R_\odot$ is caused by the crossing of the luminosity bump.


\subsection{Mass dependency of the degenerate sequence \label{sect_mass_dep}}

We now turn to the mass-dependency that was observed for the degenerate sequence in the $\dn$-$\dpun$ plane (V16). The same type of dependency is also found in stellar models with degenerate helium cores, as can be seen in Fig.\,\ref{dn_dp_deg}. We have shown that the cores of stars with similar $\Delta\Pi_1$ have nearly identical structures. The differences in $\Delta\nu$ thus come almost entirely from the envelope structure. Observations have shown that for a given value of $\dpun$, higher-mass stars tend to have larger $\Delta\nu$, and thus larger mean densities. This means that higher-mass envelopes tend to be more condensed.

\begin{figure}
\begin{center}
\includegraphics[width=9cm]{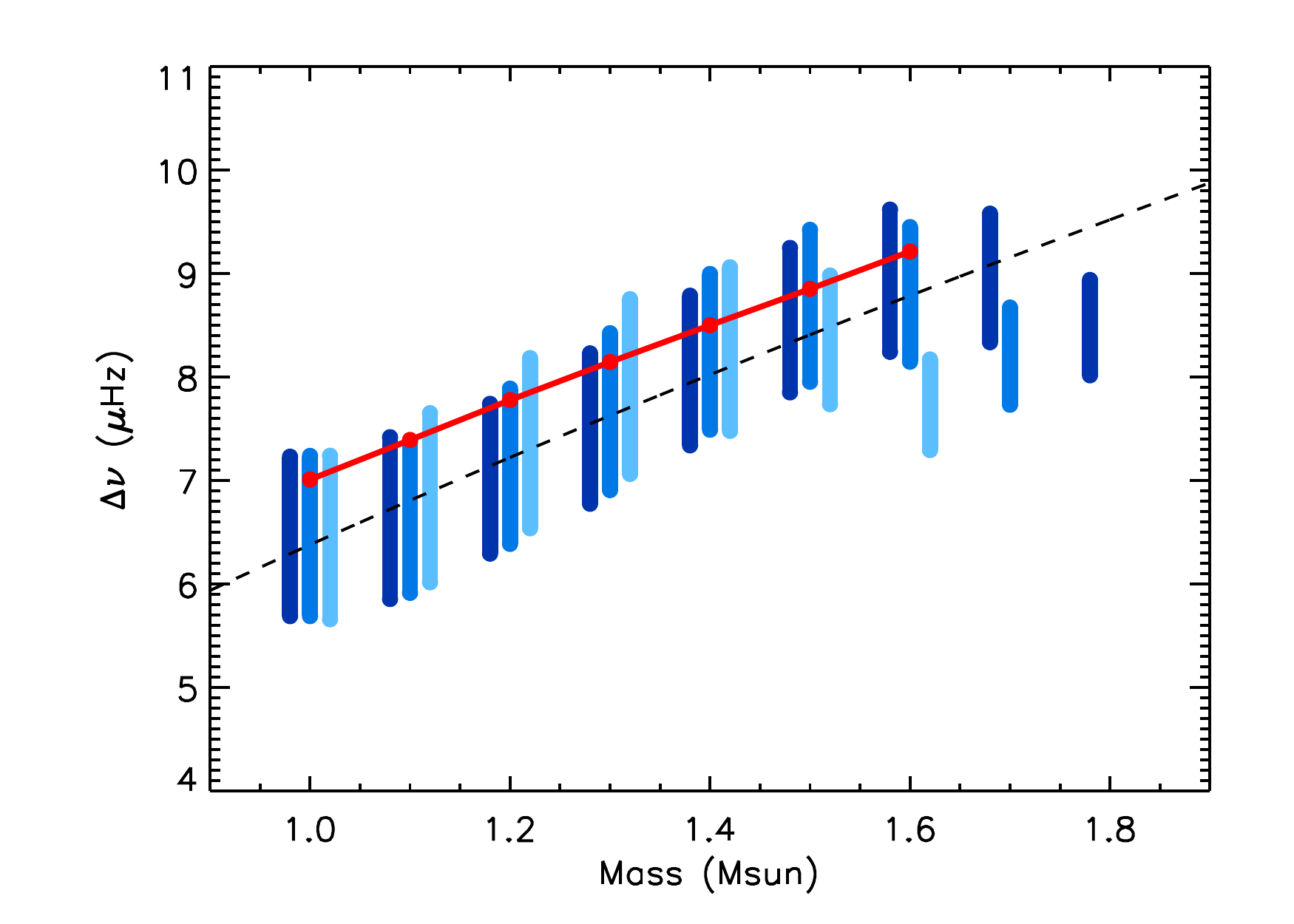}
\end{center}
\caption{Variations in $\Delta\nu$ as a function of stellar mass at fixed value of $\Delta\Pi_1$ ($71 \pm 0.5$ s). Blue filled circles represent models from our grid, with $\alpha_{\rm ov} = 0, 0.1$, and 0.2 from light blue to dark blue. For clarity, a small artificial shift in mass was applied to clearly separate models with different $\alpha_{\rm ov}$. The black dashed line corresponds to the fit of a power law to the $\Delta\nu$-$M$ relation using the models of the grid. Red circles connected by a thick red line correspond to a 1.6-$M_\odot$ stellar model to which various amounts of mass loss have been applied (see Sect.\,\ref{sect_mass_dep}).
\label{fig_dn_fixdp}}
\end{figure}

We show in Fig.\,\ref{fig_dn_fixdp} the variations in $\Delta\nu$ as a function of stellar mass for the models of our grid that have a period spacing of $\Delta\Pi_1 = 71 \pm 0.5$ s (this value was chosen arbitrarily for illustration purposes, other values yield qualitatively similar plots). In this figure, we have selected only the models that have degenerate cores. It is clear that $\Delta\nu$ increases with stellar mass at fixed $\Delta\Pi_1$. By fitting a power law to Fig.\,\ref{fig_dn_fixdp}, we find that $\Delta\nu \sim M^{0.64}$ at fixed $\Delta\Pi_1$. The value of the exponent is only weakly dependent on the chosen value of $\Delta\Pi_1$. For higher masses in Fig.\,\ref{fig_dn_fixdp}, this trend starts to reverse. Indeed, for the highest masses in the plot, electron degeneracy is not yet strong in the core and stars are just settling on the $\dn$-$\dpun$ degenerate sequence.

We then tried to provide a quantitative explanation for the mass dependency of the $\dn$-$\dpun$ degenerate sequence. At fixed period spacing $\Delta\Pi_1$, using the fact that $\dn$ scales as the square root of the mean density, we have
\begin{equation}
\left(\frac{\partial\ln\Delta\nu}{\partial\ln M}\right)_{\Delta\Pi_1} = \frac{1}{2} - \frac{3}{2} \left(\frac{\partial\ln R}{\partial\ln M}\right)_{\Delta\Pi_1}.
\label{eq_deltanu_vs_mass}
\end{equation}
As explained in Sect.\,\ref{sect_dp_R}, we expect stars with similar $\dpun$ to have the same radius approximately, that is, $\left(\partial\ln R/\partial\ln M\right)_{\Delta\Pi_1}\approx 0$. Using this crude approximation, one would have $\Delta\nu\sim M^{0.5}$, which is rather close to the relation that was obtained using the models of our grid. To refine this first estimate, one would need to take into account the mass dependency of Hayashi lines (e.g., \citealt{kippenhahn90}) and the deviations from the approximation $\left(\partial\ln L/\partial\ln M \right)_{\Delta\Pi_1} \approx 0$. The latter arises because the core structure of stars with similar $\Delta\Pi_1$ start to significantly differ near the top of the H-burning shell. Since a non-negligible part of the stellar luminosity is produced in this region, stars with common values of $\Delta\Pi_1$ can show a scatter in the total luminosity.

Instead, we followed a more pragmatic approach and proceeded to estimate $\left(\partial\ln R/\partial\ln M\right)_{\Delta\Pi_1}$ numerically. For this purpose, we evolved a 1.6-$M_\odot$ model until it reached strong electron degeneracy in the core. We then applied mass loss from the surface, in order to modify the amount of mass stored in the envelope. Since the core is degenerate, it is not affected by this procedure. This operation was repeated, applying various amounts of mass loss to create models with total masses ranging from 1 to 1.6~$M_\odot$. Using the radii of these models, we obtained $\left(\partial\ln R/\partial\ln M\right)_{\Delta\Pi_1} \approx -0.054$. This yields $\Delta\nu\sim M^{0.58}$, according to Eq. \ref{eq_deltanu_vs_mass}, which is closer to the relation that was found with our grid of stellar models.


\subsection{Mass at which core degeneracy occurs \label{sect_mass_coredeg}}

We have shown in Sect.\,\ref{sect_deg} that RGB stars with masses below about 2.1 $M_\odot$ (this limit depends on the input physics) join the $\dn$-$\dpun$ sequence on the RGB when electron degeneracy becomes strong in the core. The evolutionary stage at which these stars reach the degenerate sequence depends primarily on their mass (the lower the mass, the sooner in the evolution this occurs). Fig.\,\ref{dn_dp_deg} shows that models with the same input physics and masses of 1.7, 1.8, 1.9, and 2 $M_\odot$ reach the sequence with large separations of 14.3, 9.0, 5.4, and 2.8 $\mu$Hz, respectively. It is clear from Fig.\,\ref{dn_dp_deg} that for a given $\dn$ (that is, for a given stellar mean density), red giants below a certain mass have a degenerate core and therefore lie on the $\dn$-$\dpun$ degenerate sequence already, while stars more massive than this limit have nondegenerate cores and therefore lie above the $\dn$-$\dpun$ sequence. This transitional mass is strongly dependent on the choice of $\dn$, as is apparent from the example given above.

Another key ingredient to determine when core degeneracy becomes important is the amount of core overshooting during the main sequence. It is well known that core overshooting modifies the stellar mass that marks the transition between stars that ascend the RGB until the He flash, and stars that trigger helium burning in a nondegenerate core (e.g., \citealt{montalban13}, \citealt{bossini17}). This transitional mass indeed decreases from about 2.3 $M_\odot$ without core overshooting to about 2 $M_\odot$ for core overshooting over a distance of 0.2 $H_p$, where $H_p$ is the local pressure scale height.

\begin{figure}
\begin{center}
\includegraphics[width=9cm]{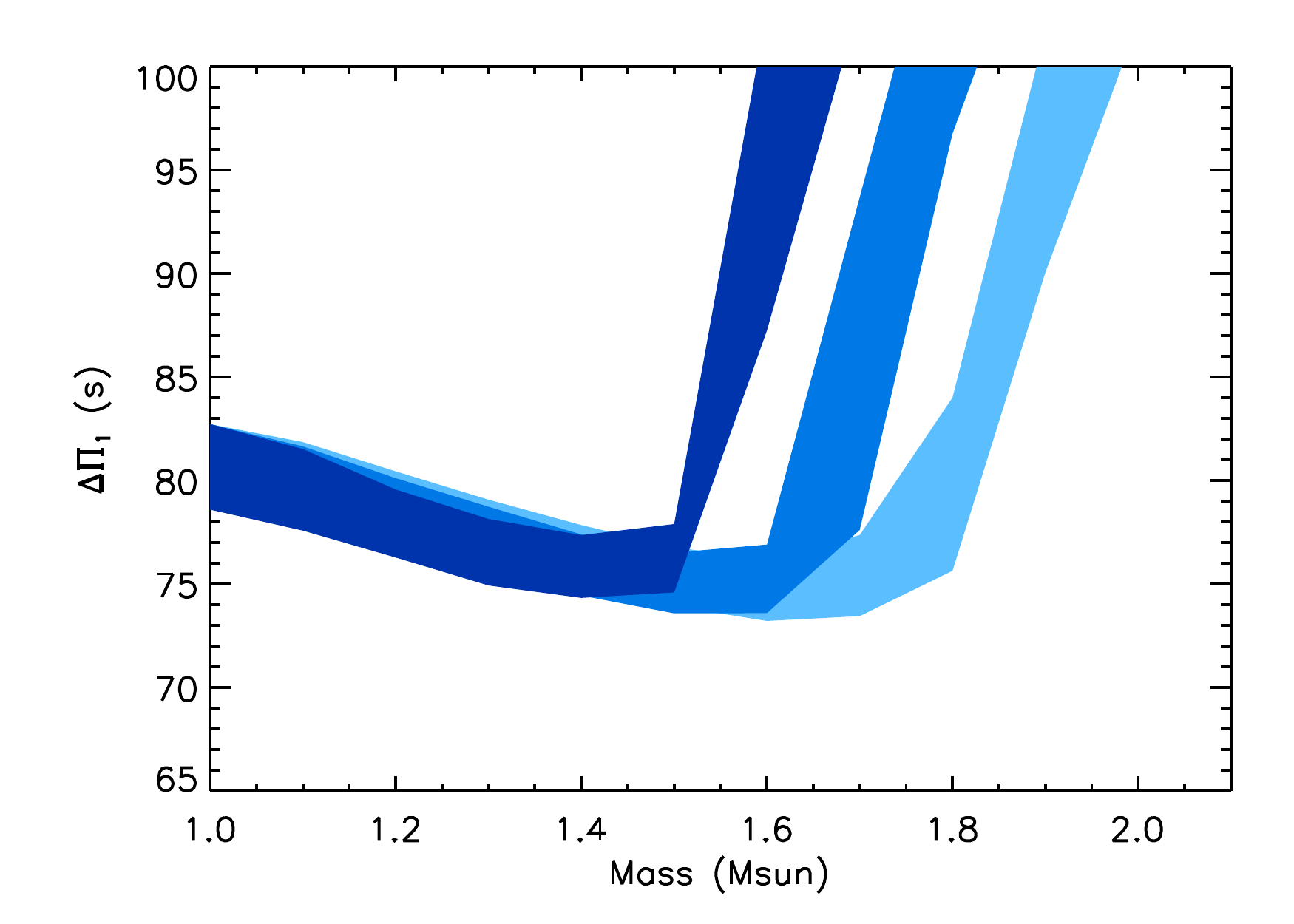}
\end{center}
\caption{Relation between $\dpun$ and stellar mass at fixed $\dn$. The shaded areas show the range of $\dpun$ for models of our grid that have a large separation of $\dn=11\pm0.5\,\mu$Hz. Core overshooting corresponds to $\aov=0, 0.1$ and 0.2 from light blue to dark blue.
\label{dp_mass_dn11}}
\end{figure}

Core overshooting during the MS also modifies the mass limit below which stars have a degenerate core at a given value of $\dn$. To illustrate this, we selected stars that have a specific large separation (here $11\pm0.5\,\mu$Hz) in the models of our grid. We plotted $\dpun$ as a function of the stellar mass for these models in Fig.\,\ref{dp_mass_dn11}, color-coding them as a function of the amount of core overshooting in the main sequence. As can be seen in the figure, the transitional mass is mainly determined by $\aov$, the metallicity having much less influence. Models without overshooting have a transition mass around 1.75 $M_\odot$, while for models computed with $\aov = 0.2$, this mass is about 1.5 $M_\odot$. A detailed comparison between models and observations for stars that lie on the degenerate sequence could therefore yield an estimate of the amount of core overshooting during the main sequence.



\section{A population of red giants below the degenerate sequence \label{sect_compare}}

\subsection{Departure from the degenerate sequence \label{sect_z}}

We mentioned in Sect.\,\ref{sect_obs} that a scatter of red giants is observed around the degenerate sequence in the $\dn$-$\dpun$ plane. 
To investigate this scatter, we first derived an approximate expression for the degenerate sequence. For this purpose, we separated RGB stars from the catalog of V16 in bins of $\dn$ of size 1 $\mu$Hz, ranging from 5 to 18 $\mu$Hz. We then calculated the median value of $\dpun$ for each bin and fit a third-order polynomial to these points, with coefficients $a_i$, $i=\{0,1,2,3\}$. This analytical approximation of the degenerate sequence is shown in Fig.\,\ref{dn_dp_vrard}. We then measured the departure from the degenerate sequence for a star with given large separation $\dn$ and period spacing $\dpun$ by the quantity $z$ defined as 
\begin{equation}
z = \dpun - \sum_{i=0}^3 a_i \dn^i.
\end{equation}
A positive (resp. negative) value of $z$ indicates a star above (resp. below) the average degenerate sequence.

\begin{figure}
\begin{center}
\includegraphics[width=9cm]{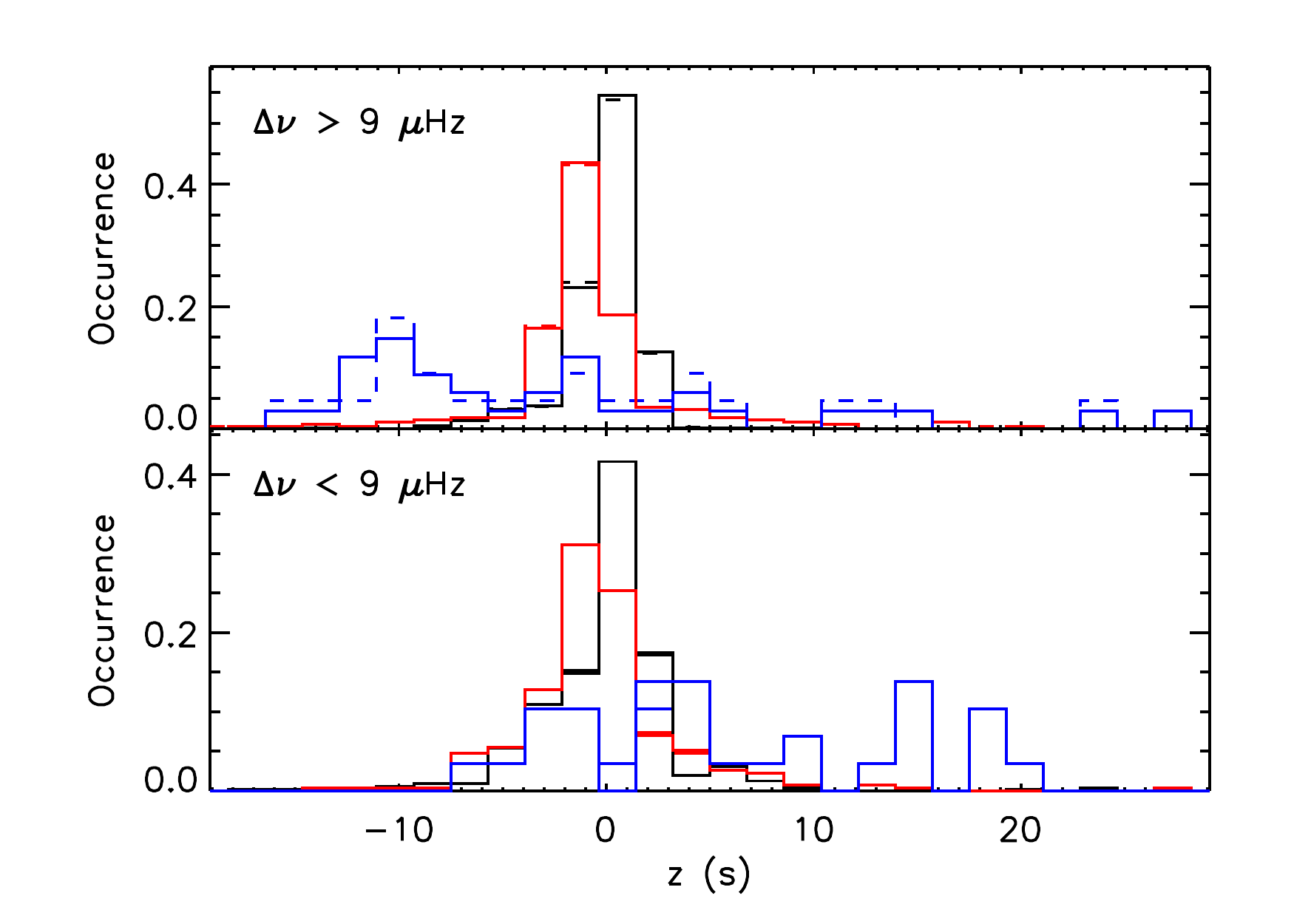}
\end{center}
\caption{Departure from degenerate sequence for low-luminosity red giants ($\dn>9\,\mu$Hz, top panel) and high-luminosity giants ($\dn<9\,\mu$Hz, bottom panel). Black, red, and blue histograms represent stars with masses $M/M_\odot<1.4$, $1.4 \leqslant M/M_\odot < 1.8$, and $M/M_\odot \geqslant 1.8$, respectively. In the top panel, the blue dashed-line histogram corresponds to stars from the catalog of V16 only, while the blue solid-line histogram also includes stars from \cite{yu18}.
\label{histo_z}}
\end{figure}

As explained in Sect.\,\ref{sect_mass_coredeg}, the stellar mass plays a critical role in determining when a star joins the degenerate sequence during its evolution on the RGB. We thus separated stars from the catalog of V16 in three mass ranges ($M/\,M_\odot < 1.4$, $1.4 \leqslant M/M_\odot < 1.8$, and $M \geqslant 1.8\,M_\odot$). We did not directly use the mass estimates that V16 derived from seismic scaling relations, but rather used the measurements of \cite{yu18}, which include a correction factor for $\Delta\nu$ calibrated with stellar models on the RGB (\citealt{sharma16}). We also distinguished between stars on the lower RGB (defined as $\dn \geqslant 9\,\mu$Hz) and stars on the upper RGB ($\dn < 9\,\mu$Hz) because they show a different behavior.

Fig.\,\ref{histo_z} shows the distribution of $z$ for stars in the three different mass ranges. For reasons that are made clear below, we show stars with large separations above and below $9\,\mu$Hz in separate panels. We observe that the great majority of stars with masses below 1.8 $M_\odot$ have low values of $|z|$ (see black and red histograms in both panels of Fig.\,\ref{histo_z}) and are thus located on the degenerate sequence, as expected. The mode of $z$ decreases with increasing stellar mass, in agreement with the mass dependency of the degenerate sequence that is predicted by stellar models (see Sect.\,\ref{sect_mass_dep}). However, we note the detection of a small fraction of stars in this mass range with large negative values of $z$ (about 3\% of these stars have values of $z$ below $-6$\,s in both ranges of $\dn$), which means that they are located well below the degenerate sequence in the $\dn$-$\dpun$ plane, in a region where we do not expect stars based on stellar models. We address the question of these stars in Sect.\,\ref{sect_below}.

For higher masses ($M/M_\odot>1.8$), the values of $z$ are much more spread out. We do expect a large proportion of these stars to have nondegenerate cores, especially for stars on the lower RGB. These stars should thus be lying above the $\dn$-$\dpun$ degenerate sequence (positive values of $z$). We indeed find a larger proportion of stars with large positive values of $z$ among higher-mass giants. However, a large fraction of higher-mass stars are found to have large \textit{negative} values of $z$ (see the peak centered around $z\sim-10$\,s in the top panel of Fig.\,\ref{histo_z}), which means that these stars are lying well \textit{below} the degenerate sequence in the $\dn$-$\dpun$ plane. These peculiar stars are mainly found among low-luminosity red giants ($\dn \geqslant 9\,\mu$Hz, top panel of Fig.\,\ref{histo_z}). For higher-luminosity giants ($\dn < 9\,\mu$Hz), this population of intermediate mass stars lying below the degenerate sequence in the $\dn$-$\dpun$ plane is much less prominent (bottom panel of Fig.\,\ref{histo_z}).

\begin{figure}
\begin{center}
\includegraphics[width=9cm]{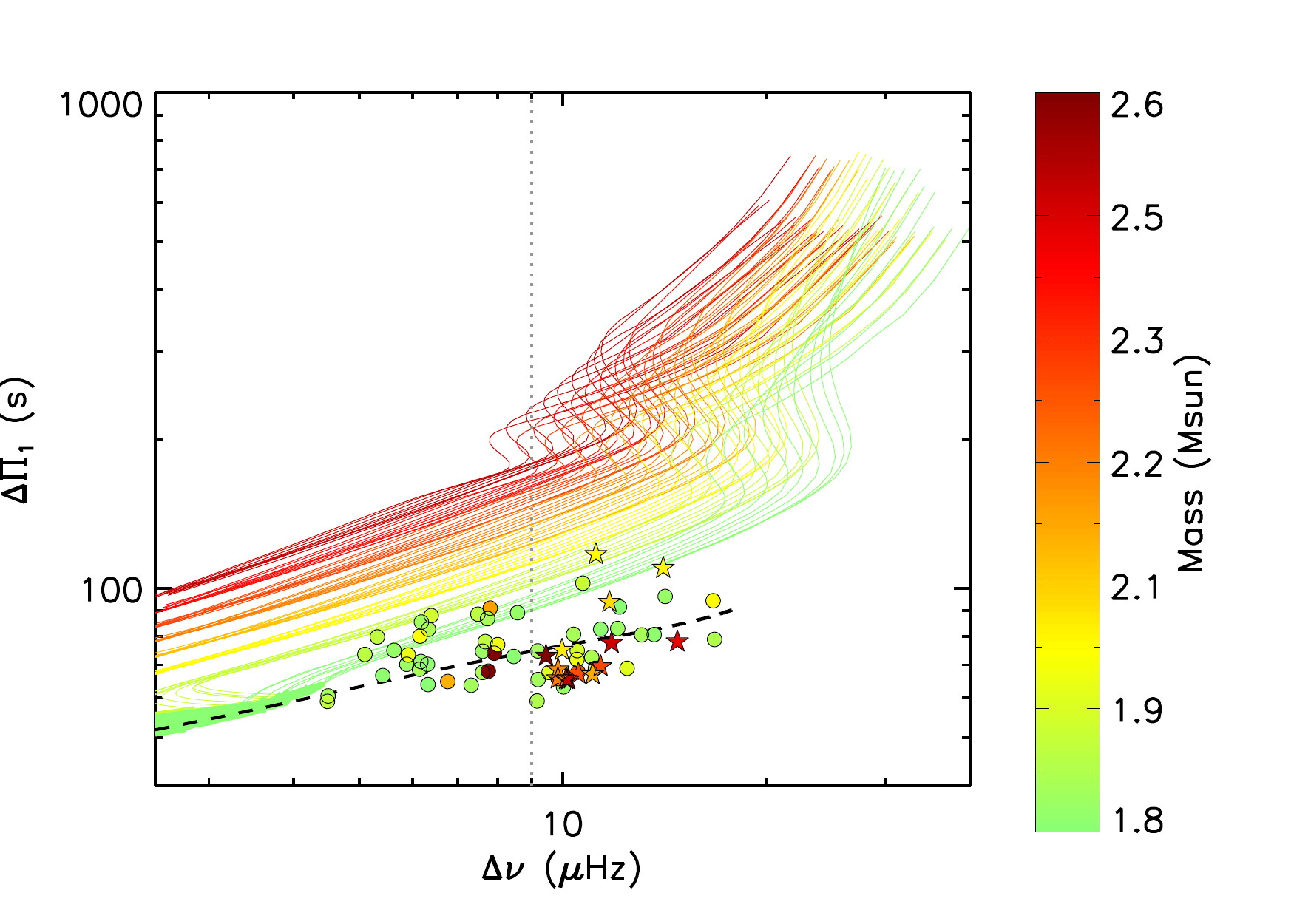} 
\end{center}
\caption{Location in the $\dn$-$\dpun$ plane of RGB stars with masses above 1.8\,$M_\odot$. The colors indicate the stellar mass, following the same scale as in Fig.\,\ref{dn_dp_deg}. Evolutionary tracks of stellar models with $\aov = 0.1$ are displayed with thick lines when electron degeneracy is strong in the core, and thin lines otherwise. Circles correspond to RGB stars studied by V16 and star-shaped symbols indicate stars from \cite{yu18} with period spacing measured in this study (see Sect.\,\ref{sect_yu18}). The black dashed line corresponds to the average degenerate sequence obtained from observations. The vertical gray dotted line shows the limit of $\dn=9\,\mu$Hz, above which we have performed new measurements of $\dpun$ in Sect. \ref{sect_below}.
\label{fig_dn_dp_mass}}
\end{figure}

This population of stars is doubly peculiar. First, stars with masses above 1.8\,$M_\odot$ are not expected to have degenerate cores in the low-luminosity part of the RGB and they should thus have period spacings that place them well above the degenerate sequence. This is illustrated in Fig.\,\ref{fig_dn_dp_mass}, which shows the evolutionary tracks of stellar models with $M>1.8\,M_\odot$ in the $\dn$-$\dpun$ plane, along with RGB stars with seismic masses above $1.8\,M_\odot$ from the catalog of V16. Clearly, a large fraction of stars observed in this mass range have period spacings that are much lower than those predicted by stellar models. To give an idea of the magnitude of this discrepancy, we can quote the example of KIC7778197, which has a seismic mass of $2.21\pm0.14\,M_\odot$, a large separation of $9.84\pm0.02\,\mu$Hz, and a measured period spacing of $\dpun = 65.7 \pm 0.55$\,s. By comparison, at the same large separation, stellar models of $2.2\,M_\odot$ all have values of $\dpun$ that are above 120\,s. This discrepancy could be caused by a wrong measurement of the stellar mass from seismic scaling relations. However, large corrections would be needed to reconcile observations with model predictions. Following on the example given above, our grid indicates that models computed with $\aov=0.1$ need to have a mass below 1.7\,$M_\odot$ in order to be located on the degenerate sequence at a large separation of $9.84\,\mu$Hz. With an overshooting efficiency of $\aov=0.2$ (more likely for this mass range), the upper mass limit drops to 1.5\,$M_\odot$. The stellar mass would therefore need to be overestimated by 23\% for $\aov=0.1$, and 32\% for $\aov=0.2$ to solve this discrepancy. This is much larger than the level of accuracy of seismic scaling relations for stellar masses, which was recently estimated to about 5\% (\citealt{serenelli21}). Another possibility to explain this discrepancy is that the period spacings might have been incorrectly estimated for these stars. To investigate this hypothesis, we check the measurements of $\dpun$ in Sect.\,\ref{sect_below}.

The second reason why these stars are peculiar is that stellar models clearly show that stars are not expected to be located below the degenerate sequence in the $\dn$-$\dpun$ plane. They should approach this sequence from above when their core is nondegenerate, and evolve along this sequence once electron degeneracy becomes strong in the core (see Fig.\,\ref{dn_dp_deg}). The lowest value of $z$ obtained for the models of our grid is around $-4.5$\,s, that is, much higher than the $z$ values obtained for a large fraction of low-luminosity giants with masses above 1.8\,$M_\odot$. We note that even if the masses of these stars were incorrectly determined, as suggested above, this would not explain why they are found below the degenerate sequence in the $\dn$-$\dpun$ plane.

\subsection{Checking the measurements of $\dpun$ \label{sect_below}}

One obvious hypothesis to explain the discrepancy that was found in the previous section would be that the asymptotic period spacings have been incorrectly measured. To measure $\dpun$, V16 perform a stretching of the oscillation spectrum in order to reproduce the evenly-spaced period pattern of pure gravity modes (see \citealt{mosser15}). For this purpose, an initial guess for $\dpun$ is required. Then, to search for regularities, a Fourier transform of the stretched spectrum is computed, which yields a revised estimate of $\dpun$. The final value of $\dpun$ is obtained by iterating on this procedure. One caveat is that the amplitudes of g-dominated mixed modes are much smaller than those of p-dominated modes, owing to their higher inertia. This produces a window effect on the stretched spectrum, and thus generates aliases, which can be mistaken for the asymptotic period spacing $\dpun$. V16 flagged the stars that might suffer from this aliasing effect in their catalog. We note that the iterative procedure followed by the authors might amplify the aliasing phenomenon. Indeed, if an alias is chosen at the first iteration, the stretching of the oscillation spectrum in the next iterations is performed with an incorrect value of $\dpun$, which favors the eventual detection of the alias. 

To investigate the reliability of the $\dpun$ estimates, we performed our own measurements using the method described in \cite{deheuvels15}, which consists in fitting an asymptotic expression of mixed mode frequencies to the detected peaks in the oscillation spectrum. We included the effects of rotation to the mixed mode frequencies in order to be able to treat stars for which the identification of rotational multiplets is not straightforward. Similar tests were performed with the method exposed in \cite{mosser18}, which also considers rotational multiplets. Details of our method are recalled in Appendix \ref{app_deltapi}. This method is less prone to the aliasing problem for two reasons. First, all possible values of $\dpun$ are consistently tested, so the method does not depend on the choice of an initial guess. Secondly, to correctly match the detected peaks, a solution from our method has to reproduce the whole pattern of rotational multiplets, which can help to distinguish between potential aliases.

\begin{figure*}
\begin{center}
\includegraphics[width=9cm]{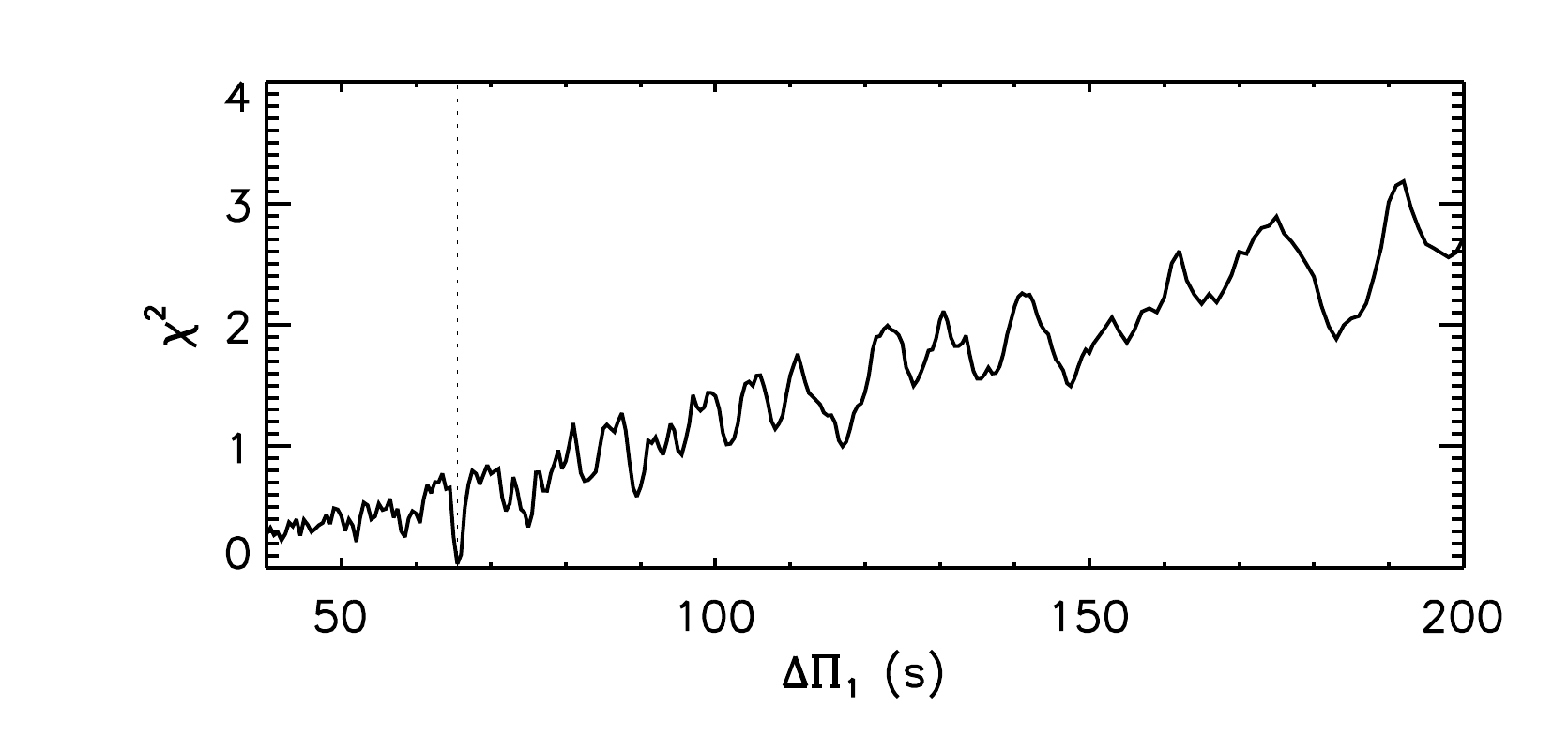}
\includegraphics[width=9cm]{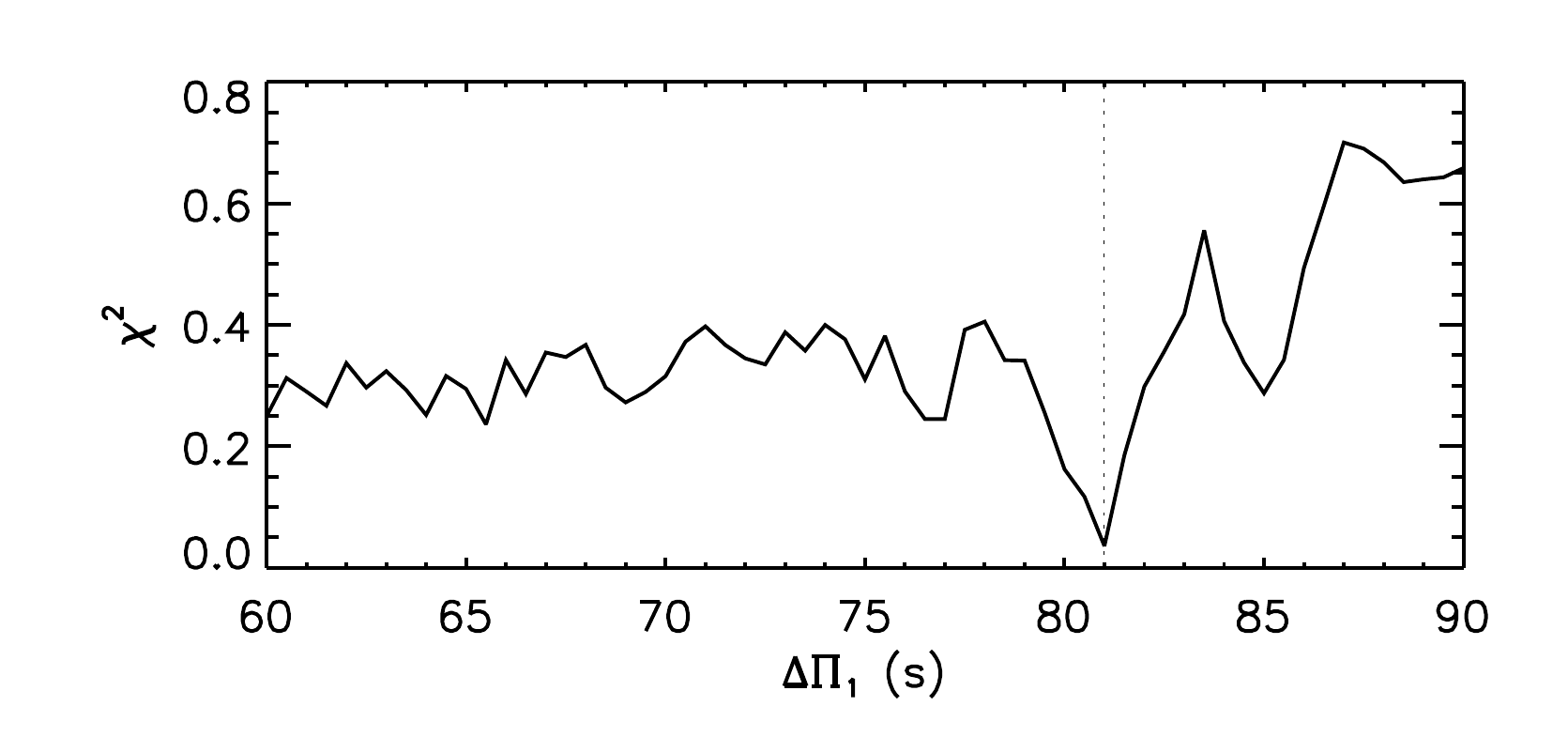}
\includegraphics[width=9cm]{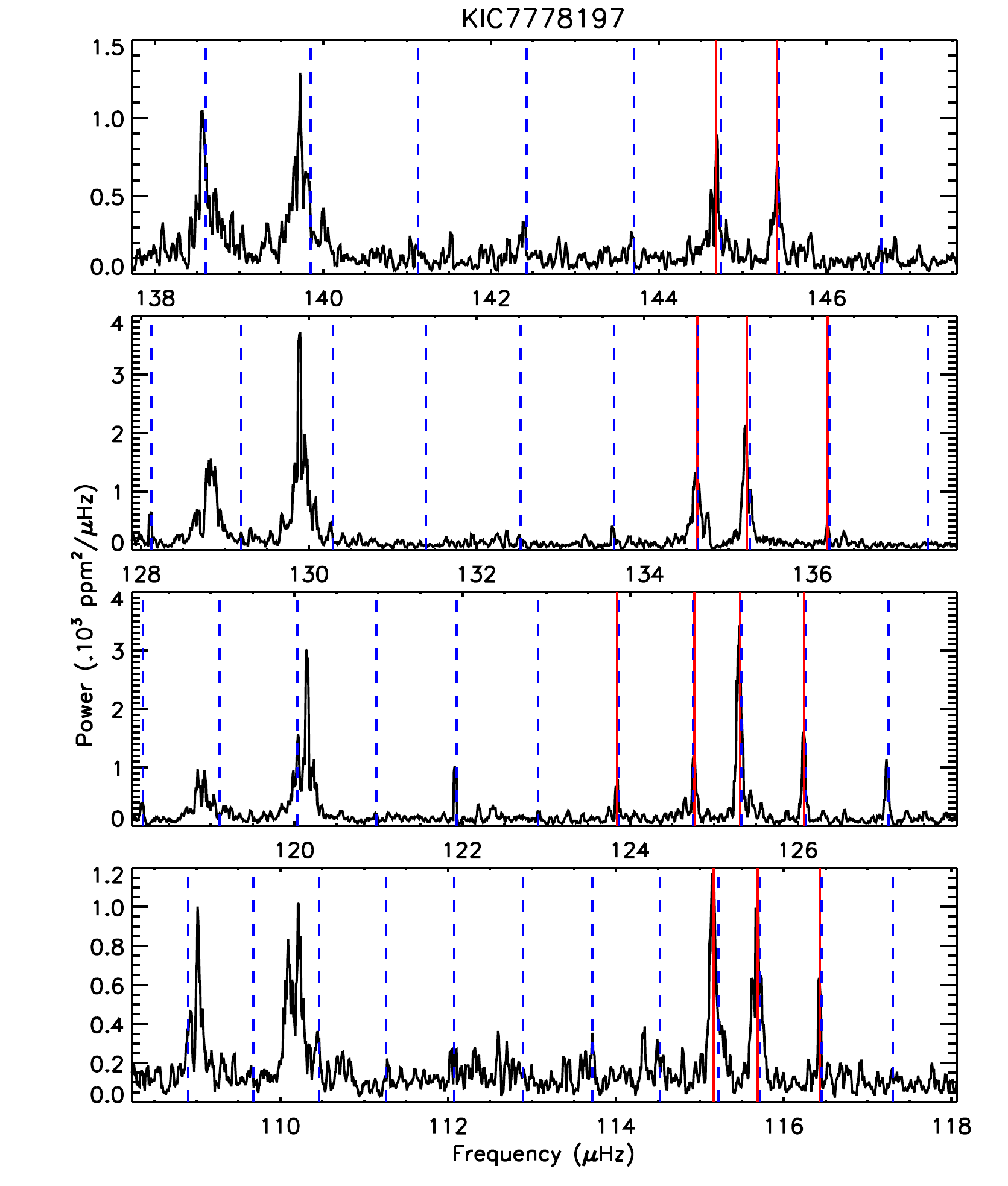}
\includegraphics[width=9cm]{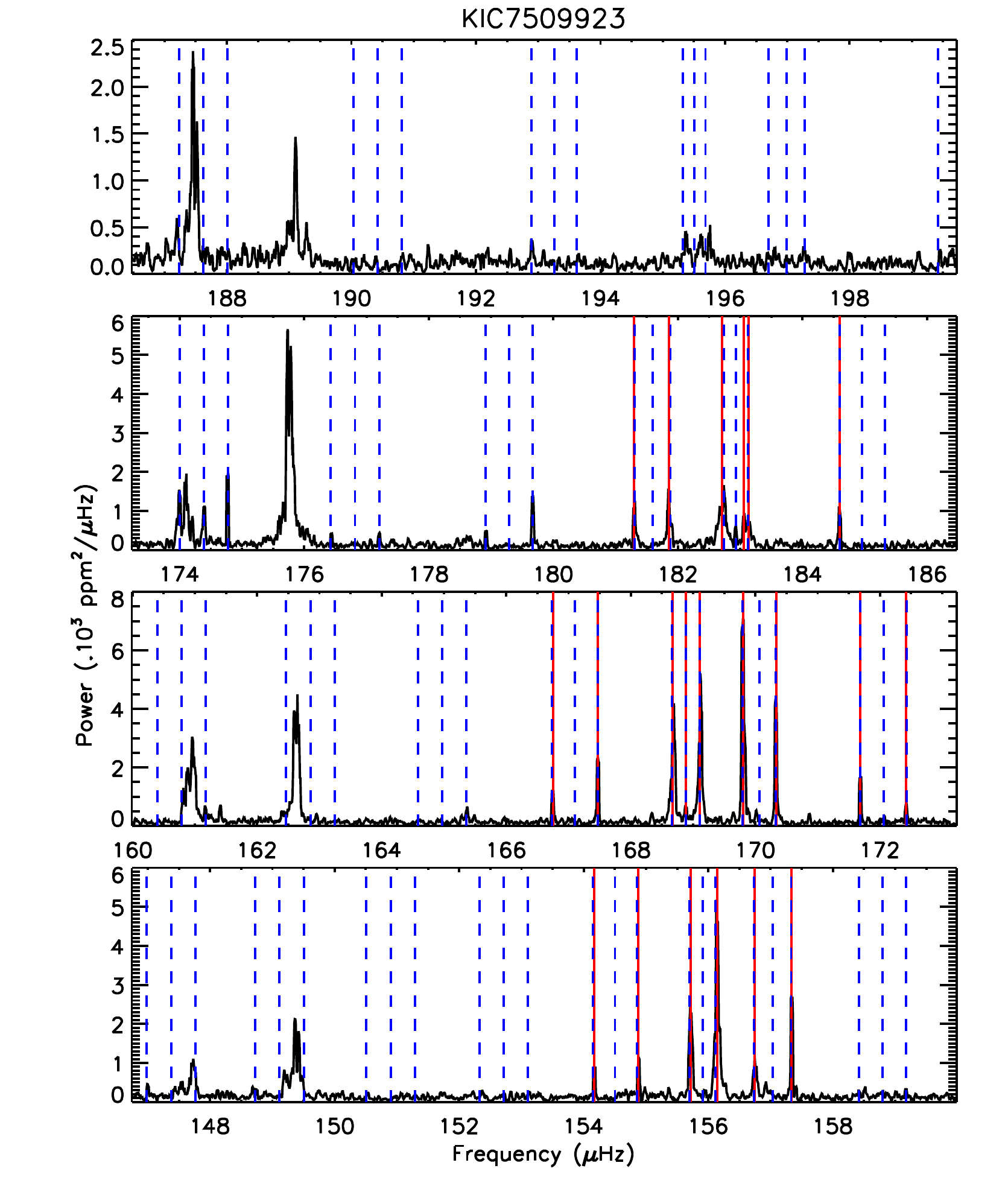}
\end{center}
\caption{Estimate of $\dpun$ for KIC7778197 (left panels) and KIC7509923 (right panels) using the method presented in Sect.\,\ref{sect_below} and Appendix \ref{app_deltapi}. \textit{Top panel:} Optimal values of $\chi^2$ (see Eq. \ref{eq_chi2}) obtained for each tested value of $\dpun$. The vertical dotted line indicates the best fit to the data. \textit{Bottom panel:} Oscillation spectrum shown in the shape of an \'echelle diagram. Blue vertical dashed lines correspond to the best fit to the detected peaks, which are indicated as red vertical solid lines. No clear signature of rotation was found in the spectrum of KIC7778197, whereas it is evident for KIC7509923 (see text).
\label{fig_7778197}}
\end{figure*}

We here show two examples of the results obtained with our method for stars that were found below the degenerate sequence.
\begin{itemize}
\item \textit{KIC7778197:} V16 found a period spacing of $\dpun = 65.7\pm0.55$\,s ($z = -10.6$\,s) for this star, but flagged the measurement as potentially corresponding to an alias. We show in the left panel of Fig.\,\ref{fig_7778197} the result of our method applied to this star. For each considered value of $\dpun$, we give the best value of $\chi^2$ (see Eq. \ref{eq_chi2}) obtained with our method. The result is here unambiguous, and we confirm the estimate of $\dpun$ from V16. Only one component per multiplet is visible for this star, which can either mean that the core rotation is slow, or that the inclination angle is low. When trying to fit the detected modes with two or three visible components per multiplet, no satisfactory solution can be found.
\item \textit{KIC7509923:} V16 found $\dpun = 65.9\pm1.15$\,s, which places this star well below the degenerate sequence ($z = -16$\,s), but they also flagged it as likely to be affected by aliasing. By applying the method of V16 to this star, we indeed recovered the solution given by the authors, along with a clear alias for $\dpun \sim 80$\,s. Comparatively, our method provided an unambiguous measurement of $\dpun = 80.9$\,s for this star, as is shown in the right panel of Fig.\,\ref{fig_7778197}. Three components per multiplet are visible for this star, and we could thus estimate the core rotation rate to $\omg/(2\pi) = 800$\,nHz. Our estimates of $\dpun$ and $\omg$ are consistent with those of \cite{gehan18}, who also measured the core rotation of this star. The solution favored by V16, around $\dpun = 65.9$\,s,  yields no satisfactory solution with our method because it cannot match the whole pattern of the rotational multiplets. This example illustrates how our method can alleviate the problem of aliasing. 
\end{itemize}

We then proceeded to check the determination of $\dpun$ for all the red giants that were found significantly below the degenerate sequence in the $\dn$-$\dpun$ plane. Taking into account the predictions of stellar models, which give a lower possible value of $z$ of about $-4.5$\,s, and the uncertainties in the measurement of $\dpun$, which are typically of the order of 1\,s, we considered that red giants are below the degenerate sequence whenever their value of $z$ is below $-6$\,s. Since most of the red giants identified as peculiar in Sect.\,\ref{sect_z} have low luminosities, we restricted our analysis to stars on the lower RGB, with large separations above 9 $\mu$Hz. We thus obtained a sample of 42 stars that satisfy these criteria, with masses ranging from 1.03 to $2.28\,M_\odot$. 

\begin{figure}
\begin{center}
\includegraphics[width=9cm]{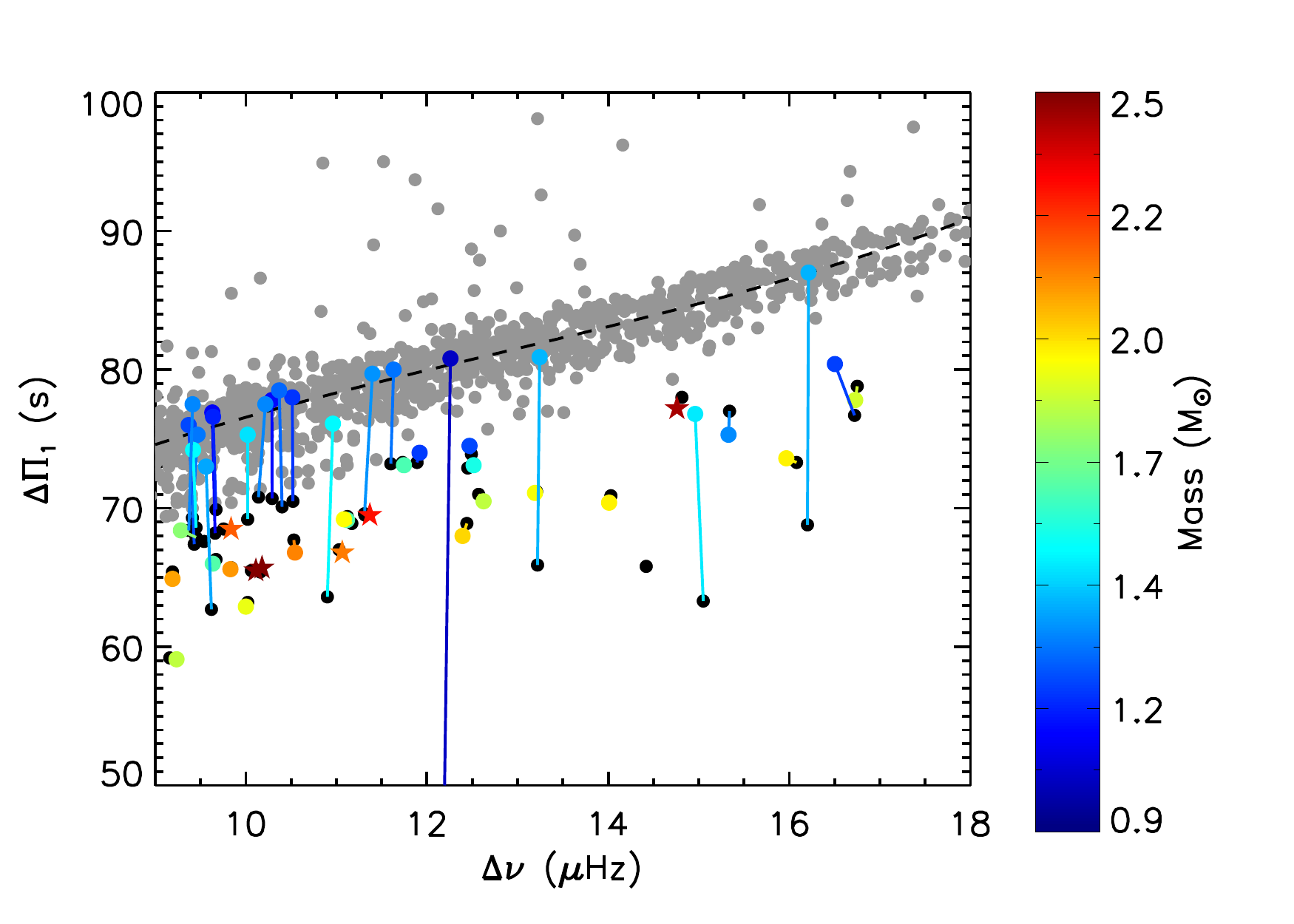}
\end{center}
\caption{Revised values of $\dn$ and $\dpun$ for the stars that were found below the degenerate sequence ($z<-6$ s) in Sect.\,\ref{sect_z}, which are indicated by black circles. Other stars from V16 are shown as gray cricles. The black dashed line corresponds to the average degenerate sequence computed in Sect.\,\ref{sect_z}. Colored circles correspond to revisions of $\dn$ and $\dpun$ obtained in this study. They are linked to the original measurements of V16 by straight lines. Colored star-shaped symbols correspond to stars from \cite{yu18} (see Sect.\,\ref{sect_yu18}).
\label{fig_dn_dp_below}}
\end{figure}

We applied our method to all the stars of this sample. We were able to obtain estimates of $\dpun$ for 40 stars out of 42. The method failed for KIC7767409 because it has low-amplitude dipolar modes (the existence of a category of red giants with low-amplitude dipolar mixed modes is well documented, see e.g., \citealt{mosser12c}, \citealt{garcia14}), and KIC4587050, which has too few detected dipolar modes to reliably extract the period spacing using our method. Our results are detailed in Table \ref{tab_deltapi} and they are shown in Fig.\,\ref{fig_dn_dp_below}. It is striking to see that our revised measurements of $\dpun$ have a strong mass dependency. As shown by Fig.\,\ref{fig_dn_dp_below}, for nearly all the lower-mass stars of the sample, we obtained estimates of $\dpun$ that deviate from those of V16. We note that all these stars were in fact flagged by the authors as potentially resulting from aliases. Interstingly, our revised measurements bring most of these stars close to the degenerate sequence, where they are expected to be found. On the contrary, for all the stars with masses larger than 1.62 $M_\odot$, we obtained measurements of $\dpun$ that are fully consistent with the measurements of V16. We can thus confirm that these stars are located below the degenerate sequence, in contradiction with the predictions of stellar models.


\begin{table}
\begin{center}
\caption{Measurements of $\dpun$ for low-luminosity giants that were found below the degenerate sequence \label{tab_deltapi}}
\begin{tabular}{l c c c c c}
\hline \hline
\T KIC Id & $M$ & $\dn$ & $\dpun$ & $z$ & $\omg/(2\pi)$ \\
\B & $(M_\odot)$ & ($\mu$Hz) & (s) & (s) & (nHz) \\
\hline
\T 3216736 & 1.23 &   12.4 & 74.5 & $-6.1$ & 0 \\
3097360$^a$ &   2.30 &     11.3 &     69.5 & $  -9.3$ &     0 \\
3648674 &   1.31 &      9.4 &     76.0 & $   0.6$ &  1020 \\
4350501 &   1.55 &     11.1 &     69.2 & $  -9.3$ &     0 \\
4667909 &   1.25 &     11.5 &     80.0 & $   0.8$ &   820 \\
4820412 &   1.04 &     12.1 &     80.8 & $   0.6$ &   690 \\
5166068 &   1.03 &      9.6 &     76.9 & $   1.0$ &   690 \\
5180345 &   1.17 &      11.8 &     74.0 & $   -5.7$ &   0 \\
5556743 &   1.82 &     10.0 &     62.9 & $ -13.6$ &   180 \\
5639438 &   1.68 &     13.1 &     71.1 & $ -10.6$ &   160 \\
5696938 &   1.86 &      9.1 &     59.1 & $ -15.8$ &     0 \\
6024654 &   1.86 &     16.8 &     77.8 & $ -10.3$ &     0 \\
6143256 &   1.32 &     10.0 &     75.3 & $  -1.3$ &   590 \\
6182668 &   1.24 &     16.5 &     80.4 & $  -7.1$ &     0 \\
6273090 &   1.70 &     16.1 &     73.6 & $ -13.1$ &   800 \\
6620586 &   1.94 &     12.5 &     68.0 & $ -12.8$ &   320 \\
6633766 &   1.46 &     11.8 &     73.1 & $  -6.5$ &   200 \\
6690139 &   1.54 &      9.6 &     66.0 & $  -9.9$ &   340 \\
6864132 &   1.88 &     11.1 &     69.2 & $  -9.3$ &   170 \\
7376259 &   1.25 &      9.4 &     75.3 & $  -0.1$ &   790 \\
7509923 &   1.46 &     13.2 &     80.9 & $  -0.9$ &   800 \\
7778197 &   2.21 &      9.8 &     65.6 & $ -10.7$ &     0 \\
8023523$^a$ &   2.60 &     10.2 &     65.7 & $ -11.2$ &   200 \\
8055108 &   1.84 &      9.2 &     64.9 & $ -10.1$ &   290 \\
8096716 &   1.16 &     10.6 &     78.0 & $   0.4$ &   690 \\
8108732 &   1.12 &     10.2 &     77.8 & $   0.8$ &  1030 \\
8301083 &   1.10 &      9.6 &     76.6 & $   0.7$ &   750 \\
8827808$^a$ &   2.45 &     14.8 &     77.2 & $  -7.2$ &   880 \\
9227589 &   1.23 &     15.2 &     75.3 & $  -9.9$ &  1900 \\
9326896 &   1.65 &     12.4 &     73.1 & $  -7.4$ &   470 \\
9474201 &   1.53 &     15.0 &     76.8 & $  -7.9$ &     0 \\
9512519 &   1.27 &      9.6 &     73.0 & $  -2.8$ &   730 \\
9642805 &   1.63 &     12.5 &     70.5 & $ -10.3$ &     0 \\
9903598$^a$ &   2.13 &     11.0 &     66.8 & $ -11.6$ &     0 \\
9907511 &   2.28 &     10.5 &     66.8 & $ -10.6$ &   190 \\
9945389 &   1.47 &     16.2 &     87.0 & $   0.1$ &   800 \\
10132814 &   1.62 &      9.4 &     74.2 & $  -1.2$ &   460 \\
10420335 &   1.17 &     10.3 &     78.5 & $   1.3$ &   540 \\
10645209 &   1.73 &     14.0 &     70.4 & $ -12.6$ &   820 \\
10712314 &   1.69 &      9.4 &     68.4 & $  -7.1$ &   660 \\
11186389 &   1.38 &     10.1 &     77.5 & $   0.7$ &   910 \\
11302371$^a$ &   2.51 &     10.1 &     65.5 & $ -11.2$ &     0 \\
11304067 &   1.08 &      9.3 &     77.5 & $   2.2$ &   630 \\
11555266 &   1.28 &     11.2 &     79.7 & $   1.0$ &   680 \\
11805876$^a$ &   2.18 &      9.8 &     68.5 & $  -7.6$ &     0 \\
\B 12301741 &   1.56 &     10.9 &     76.1 & $  -2.0$ &   730 \\
\hline
\end{tabular}
\end{center}
\small{$^a$: Stars taken from the catalog of \cite{yu18}}
\end{table}

\subsection{Intermediate-mass RGB stars from \cite{yu18} \label{sect_yu18}}

To further investigate this puzzling discrepancy, we included to our data set intermediate-mass stars from the catalog of \cite{yu18}. This catalog does not include measurements of the asymptotic period spacing, so the evolutionary status of stars cannot be known upfront. We decided to consider only low-luminosity giants ($\Delta\nu>9\,\mu$Hz) from \cite{yu18} in order to ensure that our sample only contains RGB stars (secondary clump stars have large separations below $9\,\mu$Hz, see \citealt{mosser14}). We found 43 stars with masses above 2 $M_\odot$ that satisfy this criterion and are not already within the sample of V16 (their masses range from 2.0 to 3.1 $M_\odot$).

To estimate $\dpun$ for these stars, we used the method described in Sect.\,\ref{sect_below} and Appendix \ref{app_deltapi}. We were able to measure $\dpun$ for 12 stars out of the 43 that were selected. Among the remaining stars of the sample, 21 were found to have suppressed dipolar modes, so that we could not measure $\dpun$. This number is consistent with the estimate of \cite{stello16}, who found that approximately 40\% of stars with masses above 2 $M_\odot$ show suppression of dipolar mixed modes. Nine other stars have either a too low signal-to-noise ratio or too few dipolar modes detected to produce a reliable estimate of $\dpun$. For the last star, KIC7592662, the mode parameters were incorrectly determined by \cite{yu18} because the whole oscillation spectrum actually corresponds to aliasing, the mode frequencies being higher than the Nyquist frequency for long-cadence data. The frequency of maximal power of the oscillations is $\nu_{\rm max} = 351.3 \pm 2.5\, \mu$Hz instead of $214.6\pm 2.5\, \mu$Hz as was initially found by \cite{yu18}, and the large separation is $\Delta\nu = 23.8\pm0.3\, \mu$Hz instead of $14.22\pm0.06\,\mu$Hz. With these revised parameters, this star has a seismic mass of $1.33\pm0.12\,M_\odot$ and thus does not belong to our sample of intermediate mass stars.

We added the 12 intermediate mass stars for which we could measure $\dpun$ to the bottom right panel of Fig.\,\ref{fig_dn_dp_mass} (red star-shaped symbols). Interestingly, nine out of these 12 stars are found to be located near or below the degenerate sequence, in striking disagreement with the predictions of stellar models.

\subsection{Characteristics of stars below the degenerate sequence}

To summarize, we here give the characteristics of the population of low-luminosity red giants that are found to be located below the degenerate sequence. First, they are significantly more massive than regular red giants. The average mass of red giants with confirmed values of $z$ below $-6$\,s within the catalog by V16 is 1.69\,$M_\odot$. If we include the stars from \cite{yu18}, it increases to 1.84\,$M_\odot$. By comparison, using the data set from \cite{yu18}, we find that the average mass of RGB stars with $\dn>9\,\mu$Hz is 1.34\,$M_\odot$. 

\begin{figure*}
\begin{center}
\includegraphics[width=16cm]{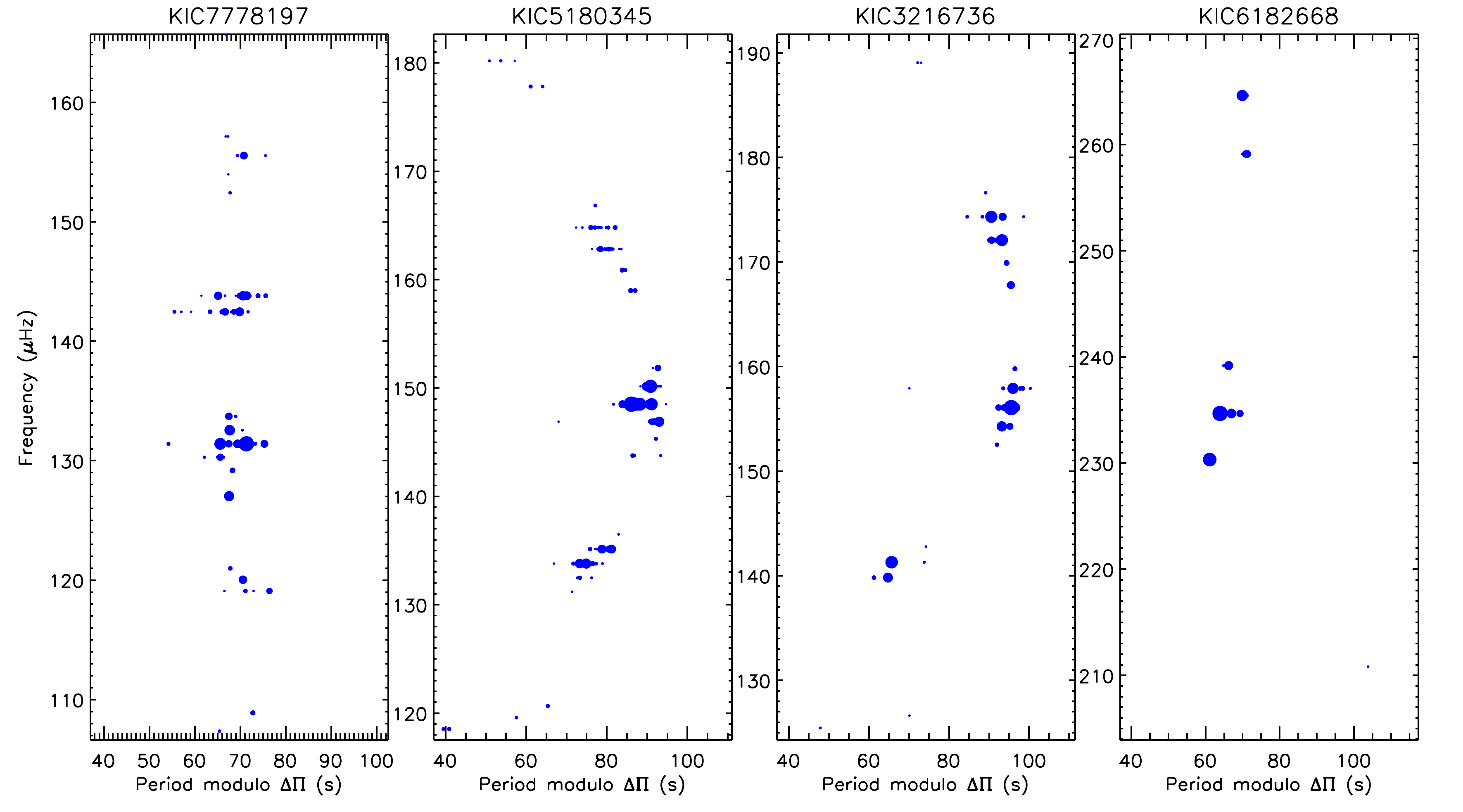}
\end{center}
\caption{Stretched period \'echelle diagrams of four targets that were found to be located below the degenerate sequence in the $\dn$-$\dpun$ plane.
\label{fig_stretch}}
\end{figure*}

We also find a few lower-mass stars below the degenerate sequence. However, these stars seem to be rather peculiar. We here show the example of KIC5180345, which has a stellar mass of $1.17\pm0.07\,M_\odot$. Our method provides an estimate of $\dpun$ around 74\,s for this star and only one component per multiplet is visible. However, this solution yields a rather poor fit to the data. This can be better understood by drawing the \'echelle diagram of stretched periods for this star, following \cite{mosser15}. When only the $m=0$ component is visible, we expect to see a nearly straight vertical ridge in the stretched \'echelle diagram (see the left panel of Fig.\,\ref{fig_stretch}). For KIC5180345, the ridge shows a clear curvature. This is reminiscent of the behavior that is expected in the presence of so-called buoyancy glitches, that is, when the \vaisala\ frequency profile shows sharp spatial variations over length scales that are shorter than the mode wavelength (see \citealt{mosser15}). We observe a similar behavior for two other low-mass RGB stars that were found to be below the degenerate sequence: KIC3216736 ($1.23\pm0.08\,M_\odot$), and to a lesser extent KIC6182668 ($1.24\pm0.09\,M_\odot$), as can be seen in Fig.\,\ref{fig_stretch}. Buoyancy glitches are not expected for RGB stars, but in core-helium burning stars only (\citealt{cunha19}). So far, the signature of a buoyancy glitch was detected in only one RGB star (\citealt{mosser18}), namely KIC3216736, which is also part of our sample. This suggests that there might be a connexion between the existence of a glitch and the fact that the period spacing is lower than expected. This clearly deserves further investigation, although this is beyond the scope of the present study. We note that the higher-mass RGB stars that are below the degenerate sequence do not show this behavior in the stretched \'echelle diagram (see for instance the case of KIC7778197 in the left panel of Fig.\,\ref{fig_stretch}).

We give in Table \ref{tab_deltapi} the core rotation rates that we obtained for the RGB stars that are located below the degenerate sequence. Among the 14 stars with masses above 1.8\,$M_\odot$ that we analyzed, seven stars show no sign of rotational splittings (only one component is detected per rotational multiplet), which means that they either have very slow core rotation or that they are seen pole-on. For the seven other targets, we obtained an average core rotation rate of $\omg/(2\pi) = 319$\,nHz. By comparison, in the catalog of \cite{gehan18}, about 26\% targets show no sign of rotational splitting and for the other stars, the average core rotation rate is 694\,nHz. This suggests that the intermediate-mass stars that are found below the degenerate sequence might have slower core rotation compared to typical red giants, although the low number of targets prevents us from drawing firm conclusions.

\section{Potential evidence for mass transfer from close red giant companion \label{sect_binary}}

The existence of stars located below the degenerate sequence in the $\dn$-$\dpun$ plane is a puzzle. This is even more intriguing since these stars are mostly intermediate-mass giants, which should have nondegenerate cores and be located well above the degenerate sequence. In this section, we propose a possible scenario that could potentially account for this double peculiarity.

We first make the hypothesis that these stars have degenerate helium cores. This seems to be a reasonable assumption because they seem to evolve on a sequence that is roughly parallel to the degenerate sequence in the $\dn$-$\dpun$ plane (see Fig.\,\ref{fig_dn_dp_below}). If it is indeed the case, then the cores of these stars have an internal structure that is nearly identical to those of regular red giants having similar values of $\dpun$, as mentioned in Sect.\,\ref{sect_rhoc}. Only their envelopes (the layers above the H-burning shell) are different. Their large separations $\dn$ are larger than those of regular red giants on the degenerate sequence. This means that they have higher mean densities, and thus that their envelopes are denser.

We have shown in Sect.\,\ref{sect_mass_dep} that higher-mass envelopes tend to be more condensed (see Fig.\,\ref{fig_dn_fixdp}). This suggests that the observed stars might result from an episode of mass accretion in the envelope of a low-mass RGB star. Indeed, let us consider a red giant with an initial mass such that it has a degenerate core on the lower part of the RGB. This star lies on the degenerate sequence. If at some point this star experiences mass accretion, for instance from a close stellar companion, its envelope will gain mass and therefore become denser. On the other hand, the structure of the degenerate core will remain unchanged because it is independent of the evolution of the envelope. The star will therefore keep a constant value of $\dpun$ but its large separation will increase. It will thus move to the right in the $\dn$-$\dpun$ plane and be located below the degenerate sequence. Interestingly, the hypothesis of mass accretion is able to account for both peculiarities observed in Sect.\,\ref{sect_compare}: it creates intermediate-mass stars with degenerate cores and it forces these stars to move below the degenerate sequence in the $\dn$-$\dpun$ plane.

We believe that this could be achieved through the evolution of a close binary system with appropriate properties. The initial separation between the two components needs to be large enough so that the two components remain well separated during the main-sequence evolution, and mass transfer occurs when the primary has already reached the RGB (this corresponds to the so-called ``case B'' of binary evolution scenarios, see e.g., \citealt{iben85}). The primary component evolves faster than the secondary. Due to the expansion of its radius in the RGB phase, it eventually fills its Roche lobe and starts transferring mass to the secondary star. In order to match the scenario described in the previous paragraph, we need the secondary star to have already developed a degenerate helium core by the time it starts accreting mass. For this reason, the mass ratio between the two components needs to be close to unity, so that the two stars are at a similar evolutionary stage at the same age. If the mass ratio were significantly smaller than unity, the secondary star would start accreting while it is still on the main sequence. Its core structure would then simply adjust to mass accretion and it would later evolve as a regular higher-mass star, triggering helium burning in a nondegenerate core. On the contrary, with a mass ratio close to one, the envelope of the secondary can gain mass without modifying the structure of the degenerate helium core. As the primary evolves, its radius continues to increase and eventually nearly all of its envelope is transferred to the secondary. The primary then shrinks and becomes a helium white dwarf.

\begin{figure}
\begin{center}
\includegraphics[width=9cm]{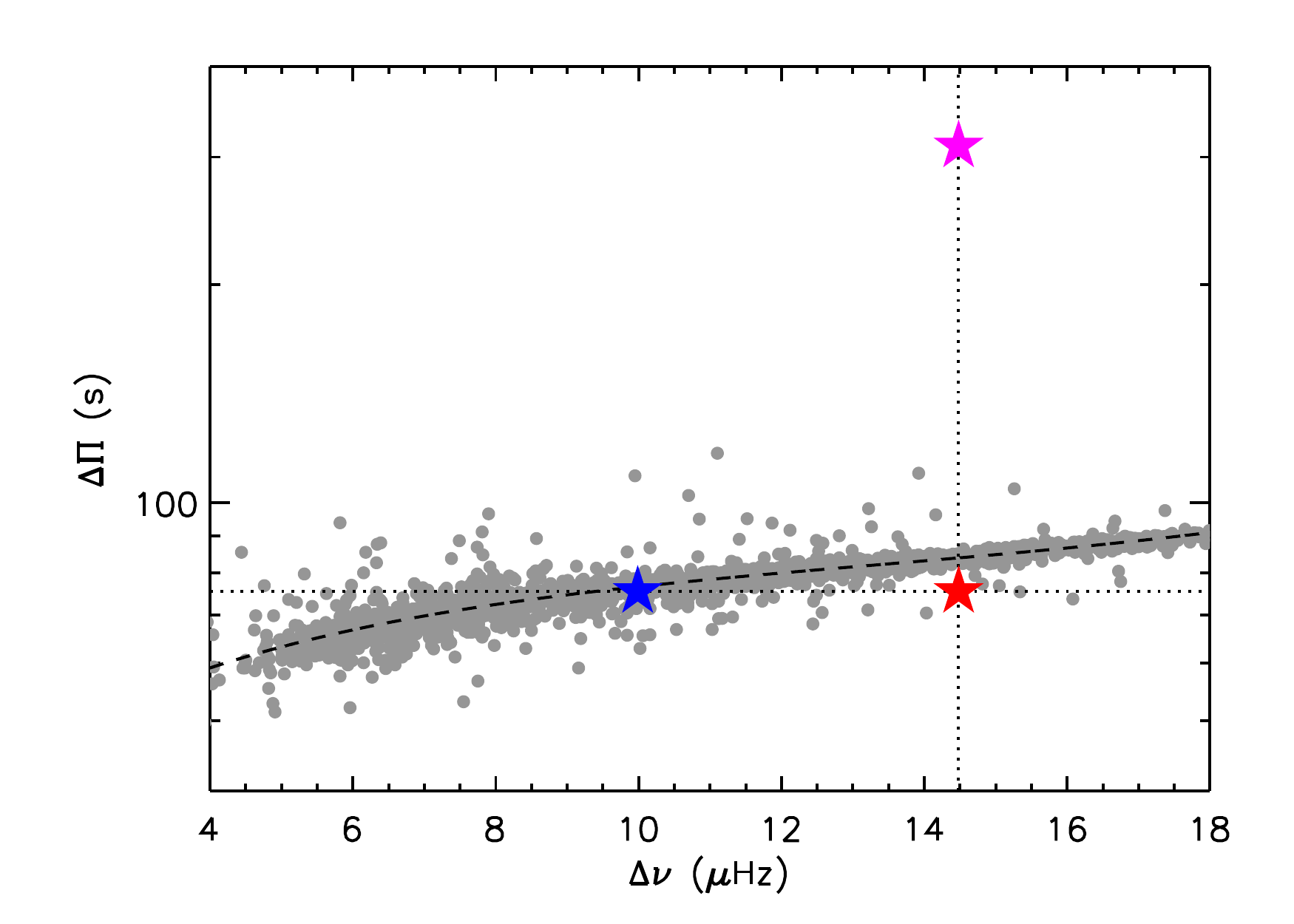}
\end{center}
\caption{Effects of mass accretion on the location of a low-mass RGB in the $\dn$-$\dpun$ plane. The blue star-shaped symbol corresponds to a regular 1.3-$M_\odot$ red giant model with solar composition. Mass accretion in the envelope of this model shifts its position horizontally to the right. The red star-shaped symbol corresponds to an accretion of 1\,$M_\odot$ (see text), which brings the stellar mass to 2.3\,$M_\odot$. The purple star-shaped symbol shows the location of a 2.3-$M_\odot$ red giant model resulting from single star evolution when it reaches the same large separation as the accretion model. The black dashed line indicates the average degenerate sequence and the gray circles correspond to RGB stars from V16, taking into account the corrections in $\dpun$ that were obtained in Sect.\,\ref{sect_below}.
\label{fig_dn_dp_massloss}}
\end{figure}

A first crude test of this scenario can be made by artificially accreting mass in the envelope of a low-mass red giant model. Let us consider for instance a binary system composed of stars with masses around 1.3\,$M_\odot$, the secondary being only slightly less massive than the primary. If the characteristics of the system are such that it evolves as described above, at the final stage of the binary evolution, the primary star will have lost nearly all of its envelope (corresponding to an overall mass loss above 1\,$M_\odot$). Assuming conservative mass transfer, the secondary will gain an equivalent amount of mass. It will then have a stellar mass above 2.3\,$M_\odot$ and a degenerate helium core (which cannot be obtained with single star evolution). To simulate the secondary, we computed a 1.3-$M_\odot$ model with solar composition using \mesa, and stopped the evolution when the asymptotic period spacing of g modes reached 75\,s (which is in the range of $\dpun$ of stars studied in Sect.\,\ref{sect_compare}). It has a large separation of 10.1\,$\mu$Hz, which places it on the degenerate sequence (blue star-shaped symbol in Fig.\,\ref{fig_dn_dp_massloss}). To simulate the effects of mass transfer, we artificially accreted 1\,$M_\odot$ to the envelope of the secondary and evolved the model to restore thermal equilibrium. As expected, this has essentially no effect on the core structure, so that the period spacing $\dpun$ is not significantly modified. Owing to the mass increase, the envelope has become denser and we now have $\dn = 14.5\,\mu$Hz. As can be seen in Fig.\,\ref{fig_dn_dp_massloss} (red star-shaped symbol), the star then lies well below the degenerate sequence in the $\dn$-$\dpun$ plane, in the vicinity of the observed stars studied in Sect.\,\ref{sect_compare}. By comparison, a 2.3-$M_\odot$ model resulting from single star evolution and evolved until it has the same large separation ($14.5\,\mu$Hz) has a nondegenerate helium core and a period spacing of 310\,s (purple star-shaped symbol in Fig.\,\ref{fig_dn_dp_massloss}).

The scenario of mass transfer from a close red-giant companion looks quite promising to explain the peculiar stars found in this study. We are currently testing this scenario more thoroughly, using binary evolution models and our results will be presented in a subsequent study. One can then wonder about the prevalence of binary systems that are in the right configuration to produce the scenario that we described. To answer this question, a more thorough study is required. However, we can already find hints using the review of \cite{moe17}. They found that about 2\% of solar-type stars ($M = 0.8$ - 1.2\,$M_\odot$) have close companions with orbital periods below 10 days (this is about the separation that is required to have mass transfer during the post-main-sequence phase) and this fraction increases rapidly with stellar mass. For low-mass close binary systems, \cite{moe17} report an excess of twins (mass ratio close to unity). This is attributed to Roche-lobe overflow during the pre-main-sequence or to shared accretion in the disk (\citealt{tokovinin00}, \citealt{halbwachs03}). \cite{moe17} found that the excess fraction of twins is about 29\% for low-mass stars, which means that about a quarter of close binary systems composed of low-mass stars have a mass ratio close to unity. Taking this figure at face value, this would mean that about 0.5\% of low-mass stars are in the right configuration to produce the scenario the we described. Among the catalog of V16, which is composed of about 2000 RGB stars, we would then expect to find about 10 targets having possibly undergone mass transfer from a close binary. Interestingly, the order of magnitude is similar to the number of targets that we identified from the catalog of V16. This requires further investigation, which will be presented in a future study.

Another possibility would be that these peculiar red giants result from the merger between a red giant and a lower-mass star. It has very recently been proposed by \cite{rui21} that stellar mergers of this type might be detectable using seismology. The underlying idea is similar to the one that we exposed in this study, that is, the accretion of a significant amount of mass in the envelope of a red giant star can produce an intermediate-mass giant with a degenerate core, which cannot be achieved with single-star evolution\footnote{The study of \cite{rui21} is completely independent of ours, and we had knowledge of this work upon submission of the present paper.}. To account for the peculiar red giants identified in the present study, this scenario would require the accreted star to be entirely transferred to the envelope of the accretor without modifying the core properties, which remains to be theoretically shown.

\section{Conclusions}

In this study, we investigated the clear sequence formed by RGB stars in the diagram showing the period spacing of dipolar g modes $\dpun$ as a function of the large separation of p modes $\dn$. Using a grid of stellar models computed with \mesa, we showed that RGB stars join this sequence when electron degeneracy becomes strong in their core. We thus propose to refer to this sequence as the degenerate sequence. The higher the mass, the later in the evolution the helium core becomes degenerate. This explains why higher-mass giants join the sequence later than their lower-mass counterparts, a feature that was pointed out from the observations (V16). Stellar models show that the degenerate sequence constitues a firm lower limit in the $\dn$-$\dpun$ diagram. Stars are thus not expected to occupy the region below this sequence.

A clear mass dependency was found for the degenerate sequence, from the observations (V16): the degenerate sequence of higher-mass stars is shifted toward lower $\dpun$. To explain this mass dependency, we first showed that red giants with degenerate cores and similar period spacings $\dpun$ have a nearly identical structure from the center up to the H-burning shell. Only their envelopes differ. Using simple physical arguments, we showed that the density of the envelope increases with increasing mass, and found that this relation approximately accounts for the mass dependency of the degenerate sequence.

The evolutionary stage at which a red giant develops a degenerate core, and thus reaches the degenerate sequence in the $\dn$-$\dpun$ plane, depends mostly on its stellar mass. However, the amount of extra mixing beyond the convective core during the main-sequence evolution also plays an important role. Stars with larger convective cores tend to reach electron degeneracy in the core earlier in the evolution. We showed that for a given value of the large separation (i.e., for a given mean density), the maximum mass of red giants that lie on the degenerate sequence can be an efficient tracer of the amount of core overshooting during the main sequence (for $\dn = 11\,\mu$Hz, this maximum mass varies from about 1.75\,$M_\odot$ without core overshooting to about 1.5\,$M_\odot$ for stars with a step overshooting over a distance of 0.2\,$H_P$). This could provide an additional measurement of the efficiency of core overshooting during the main sequence.

Within the catalog of V16, we investigated the puzzling case of a subgroup of red giants that are located below the degenerate sequence in the $\dn$-$\dpun$ plane, in contradiction with the predictions of stellar models. We performed a new measurement of the asymptotic period spacings $\dpun$ for these stars, using a different technique from V16. We showed that most of the low-mass giants in this sample are in fact located on the degenerate sequence, as expected from stellar models. However, we verified without ambiguity that the higher-mass giants of this sample are indeed located well below the degenerate sequence. This is all the more puzzling since most of these stars, owing to their high masses, should have nondegenerate cores and thus lie well above the degenerate sequence.

To account for these peculiar stars, we assumed that they have degenerate helium cores. To be located below the degenerate sequence, these stars must have higher large separations $\dn$ (and thus higher mean densities) than regular red giants with the same core density. Since the core structure is entirely determined by electron degeneracy, this means that these stars must have denser envelopes than regular red giants. We argued in this paper that this could be achieved if the envelope has gained a substantial amount of mass, for instance during an episode of mass transfer from a close companion.

We thus proposed the following scenario. We considered two low-mass stars ($\lesssim2\,M_\odot$) in a close binary system with an initial separation such that they will start to interact only during the post-main-sequence evolution. They must have a mass ratio close to unity, so that when mass transfer begins, the secondary star is already on the RGB, with a degenerate helium core. In these conditions, the envelope of the secondary can gain mass without modifying the structure of the degenerate helium core. The primary star is then expected to transfer nearly all the content of its envelope to the secondary and to eventually become a helium white dwarf. With this scenario, the secondary has gained a substantial amount of mass (for instance, it should gain more than $1\,M_\odot$ for stars with initial masses of $1.3\,M_\odot$) and has a degenerate core. Its more massive envelope is denser, so that it then has a larger $\dn$ than regular red giants on the degenerate sequence. The scenario of mass transfer from a close red giant companion is thus able to account for the two peculiarities of the sample of stars under study: the secondary star has a degenerate core despite its intermediate mass and it is lying below the degenerate sequence in the $\dn$-$\dpun$ plane. 

We are currently in the process of computing binary evolution models with \mesa\ in order to fully test the proposed scenario. Preliminary results are promising and will be presented in a following paper. The question of the probability of such a scenario will also be addressed in this follow-up work. First order-of-magnitude estimates obtained from \cite{moe17} show that the fraction of stars that should be in the right configuration to produce the scenario that we propose is roughly compatible with the number of red giants that are found below the degenerate sequence, which is quite encouraging.

\begin{acknowledgements}
We thank the anonymous referee for comments that improved the clarity of the paper. S.D. and J.B. acknowledge support from from the project BEAMING ANR-18-CE31- 0001 of the French National Research Agency (ANR) and from the Centre National d'Etudes Spatiales (CNES).
\end{acknowledgements}

\bibliographystyle{aa.bst} 
\bibliography{biblio} 

\begin{appendix}

\section{Fit of the mixed modes \label{app_deltapi}}

To determine the asymptotic period spacings of red giants, we follow the method described in \cite{deheuvels15}, which we adapt to take the effects of rotation into account. We proceed as follows.

We first estimate the global parameters of p modes by fitting the following second-order asymptotic expression to the observed radial modes:
\begin{equation}
\nu_{n,l=0} = \left[ n + \varepsilon_{\rm p} + \frac{\alpha}{2}(n-n_{\rm max})^2 \right] \Delta\nu,
\label{eq_wkb_p}
\end{equation}
where $\alpha$ measures the second-order effects in the asymptotic development, $\varepsilon_{\rm p}$ is a phase term, and $n_{\rm max} = \nu_{\rm max}/\Delta\nu$ (\citealt{mosser13}). By smoothing the power spectrum of the observed star, we obtain first approximate expressions for the frequencies of radial modes. We fit these frequencies to Eq. \ref{eq_wkb_p} and thus obtain estimates of $\Delta\nu$, $\varepsilon_{\rm p}$, and $\alpha$.

We then search for peaks in the domain where dipolar mixed modes are expected. They should have their highest amplitudes in the vicinity of $l=1$ pure pressure  modes, whose approximate frequencies can be obtained using our previous measurements along with the so-called universal pattern of \cite{mosser11}. We search for significant peaks contained within a window of width $0.45\Delta\nu$ centered on the frequencies of pure $l=1$ p modes. The width of the window is chosen in order to avoid selecting $l=2$ and $l=3$ modes. The frequencies of significant peaks are hereafter labeled as $\nu_i^{\rm obs}$, $i=1,N$.

We then use the asymptotic expression of mixed mode frequencies proposed by \cite{shibahashi79}. In this context, the matching of solutions corresponding to g-modes in the core, and to p-modes in the envelope requires that
\begin{equation}
\tan(\theta_{\rm p}) = q \tan(\theta_{\rm g})
\label{eq_wkb_mixed}
\end{equation}
where $q$ corresponds to the coupling strength between the two cavities, and $\theta_{\rm p}$, $\theta_{\rm g}$ are phase terms that can be expressed as a function of the asymptotic expressions of p- and g-modes. Following \cite{mosser12a}, we have
\begin{align}
\theta_{\rm p} & = \frac{\pi}{\Delta\nu} \left[ \nu - \left( \nu_{\rm p} \right)_{n,1} \right] \\
\theta_{\rm g} & = \pi \left( \frac{1}{\Delta\Pi_1 \nu} - \varepsilon_{\rm g} \right)
\end{align}
where $\left( \nu_{\rm p} \right)_{n,1}$ correspond to the frequencies of $l=1$ pure pressure modes. The latter can be expressed as
\begin{equation}
\left( \nu_{\rm p} \right)_{n,1} = \nu_{n,0} + (1/2 - d_{01}) \dn,
\end{equation}
where $d_{01}$ corresponds to the mean small separation built with $l=0$ and 1 pressure modes. For any given set of parameters $(\Delta\Pi_1, q, \varepsilon_{\rm g}, d_{01})$, the corresponding mixed mode frequencies $\nu_{n,1}$ can be obtained by solving Eq. \ref{eq_wkb_mixed} with a Newton-Raphson algorithm.

The effects of rotation can be considered by calculating the rotational splittings of the modes, denoted as $\delta\nu_{\rm s}$. \cite{goupil13} showed that the splitting $\delta\nu_{\rm s}$ can be expressed as a function of the average rotation rates $\omg$ and $\omp$ in the g-mode and p-mode cavities, respectively, and the parameter $\zeta$, which corresponds to the ratio between the kinetic energy of the mode in the g-mode cavity and the total kinetic energy of the mode. They obtained the relation
\begin{equation}
\delta\nu_{\rm s} = \left[ \frac{\omg/(2\pi)}{2} - \omp/(2\pi) \right] \zeta + \omp/(2\pi).
\label{eq_split}
\end{equation}
Using an asymptotic development, \cite{goupil13} found an expression relating $\zeta$ to the mode frequencies, which means that this parameter can be estimated directly from the observations. \cite{hekker17} showed that the original expression of \cite{goupil13} can be conveniently recast as
\begin{equation}
\zeta = \left[ 1 + q \frac{\nu^2\dpun}{\dn} \frac{1}{\sin^2\theta_{\rm p} + q^2 \cos^2\theta_{\rm p}} \right]^{-1}
\label{eq_zeta}
\end{equation}
Using Eq. \ref{eq_split} and \ref{eq_zeta}, we can calculate the frequencies of modes with azimuthal order $m$ as
\begin{equation}
\nu_{n,1,m} = \nu_{n,1} + m\delta\nu_{\rm s}
\end{equation}
for any given rotation profile.

We can then search for the set of parameters $(\Delta\Pi_1, q, \varepsilon_{\rm g}, d_{01},\omg,\omp)$ that produces the best match with the peaks that were detected in the expected frequency domain of $l=1$ mixed modes. In practice, the envelope rotation of red giants has been shown to have negligible contribution to the splittings of mixed modes in red giants (\citealt{mosser12b}, \citealt{goupil13}). We thus choose to neglect $\omp$ in the expression of the rotational splittings. All other parameters are varied within a grid and for each configuration, we calculate the distance between the observed peaks
\begin{equation}
\chi^2 = \sum_{i=1}^N (\nu_i^{\rm obs} - \nu_i^{\rm closest})^2,
\label{eq_chi2}
\end{equation}
where $\nu_i^{\rm closest}$ corresponds to the frequency $\nu_{n,1,m}$ that is the closest to the observed peak at a frequency $\nu_i^{\rm obs}$.

This method has already been successfully used by \cite{deheuvels15} to measure $\dpun$, but without taking rotation into account. Only axisymmetric modes were then considered in the fit, which is possible only when the identification of the azimuthal order $m$ is evident in the observations. This is not necessarily the case, especially when the rotational splitting becomes comparable to the frequency separation between mixed modes of consecutive radial orders. This is why we chose to add the effects of rotation to the expression of the modes frequencies, and to consider all detected peaks.

\end{appendix}

\end{document}